# Sample size estimation for comparing dynamic treatment regimens in a SMART: a Monte Carlo-based approach and case study with longitudinal overdispersed count outcomes

JAMIE YAP[a], JOHN J. DZIAK[b], RAJU MAITI[c], KEVIN LYNCH[d], JAMES R. MCKAY[d,e],

BIBHAS CHAKRABORTY[f,g,h], INBAL NAHUM-SHANI[a]

[a]Institute for Social Research, Ann Arbor, MI, USA

[b]Bennett Pierce Prevention Research Center, Pennsylvania State University, PA, USA

[c]Economic Research Unit, Indian Statistical Institute, Kolkata, India

[d]Department of Psychiatry, University of Pennsylvania, PA, USA

[e]Philadephia Veterans Affairs Medical Center, PA, USA

[f]Centre for Quantitative Medicine, Duke-NUS Medical School, Singapore

[g]Department of Statistics and Data Science, National University of Singapore, Singapore

[h]Department of Biostatistics and Bioinformatics, Duke University, Durham, NC, USA

**Abstract**

Dynamic treatment regimens (DTRs), also known as *treatment algorithms* or *adaptive interventions*, play an increasingly important role in many health domains. DTRs are motivated to address the unique and changing needs of individuals by delivering the type of treatment needed, when needed, while minimizing unnecessary treatment. Practically, a DTR is a sequence of decision rules that specify, for each of several points in time, how available information about the individual's status and progress should be used in practice to decide which treatment (e.g., type or intensity) to deliver. The sequential multiple assignment randomized trial (SMART) is an experimental design widely used to empirically inform the development of DTRs. Sample size planning resources for SMARTs have been developed for continuous, binary, and survival outcomes. However, an important gap exists in sample size estimation methodology for SMARTs with longitudinal count outcomes. Further, in many health domains, count data are *overdispersed* – having variance greater than their mean. We propose a Monte Carlo-based approach to sample size estimation applicable to many types of longitudinal outcomes and provide a case study with longitudinal overdispersed count outcomes. A SMART for engaging alcohol and cocaine-dependent patients in treatment is used as motivation.

## 1. Introduction

Dynamic treatment regimens (DTRs) are a powerful tool for improving health in a resource efficient manner.[1] DTRs are motivated to address the unique and changing needs of individuals by delivering the type of treatment needed, when needed, while minimizing unnecessary treatment.[2] A DTR includes a sequence of decision rules that specify, at each decision point (i.e., a point in time when treatment decisions should be made), whether and how to modify the type or intensity of treatment based on information about the individual (i.e., *tailoring variables*). DTRs are intended to provide replicable protocols for tailoring sequences of treatments in clinical practice.[3, 4]

The sequential multiple assignment randomized trial (SMART) is an experimental approach for obtaining empirical evidence necessary to construct effective DTRs.[5, 6] A SMART involves multiple stages of randomization, each stage beginning with a decision point in which some or all individuals are randomized among appropriate treatment options. As an example, consider the ENGAGE study, in which participants were 500 alcohol and cocaine-dependent individuals. The participants first completed two weeks of an Intensive Outpatient Program (IOP) — the most common treatment offered to individuals with relatively severe substance-use disorders. During this period, investigators assessed participants' engagement with the IOP by monitoring session attendance. Only those who were not adequately attending treatment (189 individuals) proceeded to the SMART where they were randomized sequentially to different strategies for re-engaging them in treatment. The goal was to investigate whether and under what conditions providing non-engaged participants opportunities to choose their preferred treatment strategy would lead to higher rates of re-engagement and to improved substance use outcomes

### 1.1 The ENGAGE SMART

Participants who proceeded to the SMART were initially randomized (Figure 1) to two types of telephone-based outreach efforts using motivational interviewing (MI)[7] principles. The first type **(MI-Choice)** highlighted the opportunity to select one's treatment option. Specifically, as part of telephone-based outreach efforts, participants were told that they could elect to switch to either cognitive-behavioral therapy, telephone-based stepped care, or medication management, or else remain in IOP. The second type **(MI-IOP)** did not highlight the opportunity to select one's preferred treatment option; instead, the telephone-based outreach efforts were solely focused on helping the individual adhere to the IOP. The goal was not to investigate the comparative effect of the treatments the individual selected or received, but rather to inform the development of a DTR that integrates the element of choice to promote treatment engagement. Nonetheless, most participants assigned to MI-Choice elected to remain in the IOP. Further details are given by McKay and colleagues.[8]

Individuals who were still not adherent at week 6 since the initial randomization (week 8 of treatment overall) were regarded as *non-responders* and re-randomized, either to **MI-Choice** or to receive no further contact by the study team **(NFC)**. Conversely, individuals who were adherent by then were regarded as *responders* and not re-randomized, henceforth receiving **NFC**. This design resulted in 6 cells, labeled A-F in Figure 1, and contains four *embedded DTRs* (EDTRs) (see Table 1).

### 1.2 Sample Size Planning Resources for SMARTs

Sample size planning resources have been developed for SMARTs having survival outcomes,[9] and for continuous (e.g., normal) or binary end-of-study outcomes.[10-12] For the continuous and binary cases, resources that allow longitudinal data to be utilized to improve sample size efficiency in comparing end-of-study outcomes already exist.[12, 13] However, sample size resources are currently not available for longitudinal data from a SMART with other kinds of outcome distributions, or for estimands focusing on the longitudinal process (e.g., area under the curve, AUC). The goal of this paper is to provide a flexible method for addressing both gaps. Specifically, we propose a Monte Carlo-based approach which integrates ideas from the potential outcome framework and copulas in the data-generating phase. This approach holds two major benefits to planning sample size for SMARTs with non-continuous longitudinal outcomes. First, it uses design parameters that are interpretable and that study designers can elicit relatively easily using existing evidence and clinical considerations. Second, it accommodates realistic features of SMARTs, including the timing of repeated measurements in relation to sequential randomizations, dependence between response status and the longitudinal outcome, and restricted randomizations. We illustrate the challenges in planning sample size for SMARTs with non-continuous longitudinal outcomes, and the utility of the proposed method, by focusing on a case study with a longitudinal overdispersed count outcome (i.e., counts with variances higher than their mean).

In ENGAGE, possible longitudinal count outcomes could be the number of therapy sessions attended in the past month or the number of alcohol-use days in the past month; both are overdispersed in the observed data. Other examples of possibly overdispersed count outcomes could include brain lesion counts in multiple sclerosis[14], counts of exacerbations in chronic obstructive pulmonary disease,[15] or counts of adverse events among cancer patients.[16] When obtaining power via simulation, it is necessary to use a realistic distribution; these counts are unlikely to resemble a normal or Poisson distribution. We demonstrate an approach that permits study designers to incorporate key features of the longitudinal outcome (e.g., overdispersion) into their sample size estimates.

### 1.3 Outline of the Manuscript

We begin by describing the inferential target (Section 2) and a Monte Carlo simulation approach to sample size estimation for SMARTs (Section 3). Throughout Section 4, we describe our proposed approach in terms of a broad framework applicable to many types of longitudinal outcomes and provide details on how the framework can be more specifically applied when the longitudinal outcome of interest is an overdispersed count. Simulation studies investigating the properties of our proposed approach are described in Sections 5 and 6. Finally, directions for future research are discussed in Section 7.

## 2. Hypothesis tests for comparing EDTRs in a SMART

We focus on one of the most common designs for SMARTs exemplified by ENGAGE (Figure 1): a *two-stage restricted SMART*. In this design, there are two possible first-stage treatment options, and two possible second-stage options; the decision on whether to randomize individuals to second-stage treatment options is determined based on a tailoring variable. Most commonly, the tailoring variable is the individual's *response status* – an indicator for whether sufficient progress was achieved during the first stage of treatment – and only individuals classified as *non-responders* are re-randomized to second-stage options.[3, 4]

## 2.1 Notation for quantities associated with EDTRs

We assume that measurement occasions are the same across all individuals. Let $N$ and $T$ denote the total number of individuals and measurement occasions, respectively, in a SMART. Let $t_j$ denote time elapsed between initial randomization and the $j^{th}$ measurement occasion; hence, time between two consecutive measurement occasions is given by $t_{j+1} - t_j$.

Denote an EDTR by $(a_1, a_2^{NR})$ where $a_1$ is a first-stage treatment option and $a_2^{NR}$ is a second-stage treatment option offered to non-responders; $a_1 \in \{+1, -1\}$ and $a_2^{NR} \in \{+1, -1\}$. Let $Y_{i,j}$ denote an outcome of an individual $i$ that was actually observed at the $j^{th}$ measurement occasion. Let $Y_{i,j}^{(a_1, a_2^{NR})}$ denote the outcome that would have been observed at the $j^{th}$ measurement occasion had individual $i$ followed EDTR $(a_1, a_2^{NR})$, i.e., the *potential outcome*[17] of individual $i$ at occasion $j$ under EDTR $(a_1, a_2^{NR})$. Similarly, let $R_i$ denote the response status (1=responder, 0=nonresponder) of an individual $i$ that was actually observed, and let $R_i^{(a_1)}$ denote response status of an individual $i$ that would have been observed had the individual undergone first-stage option $a_1$.

Finally, define the *mean trajectory* of *EDTR* $(a_1, a_2^{NR})$ as $\mathcal{T}^{(a_1, a_2^{NR})}(\boldsymbol{X}) := \left\{\mu_j^{(a_1, a_2^{NR})}(\boldsymbol{X}), j = 1, \ldots, T\right\}$ where $\mu_j^{(a_1, a_2^{NR})}(\boldsymbol{X}) \coloneqq E\left[Y_j^{(a_1, a_2^{NR})} | \boldsymbol{X}\right]$, and $\boldsymbol{X}$ denotes mean-centered baseline covariates. In words, $\mu_j^{(a_1,a_2^{NR})}(\boldsymbol{X})$ is the mean of the longitudinal outcome at the $j^{th}$ measurement occasion had all individuals with baseline characteristics $\boldsymbol{X}$ followed EDTR $(a_1, a_2^{NR})$.

### 2.2 Focusing on pairwise comparisons of EDTRs

Omnibus testing and multiple comparisons with the best are possible in SMARTs,[18] but planning sample size for a scientifically interesting pairwise comparison of two EDTRs is more common and hence the focus of this paper. The comparison of EDTRs based on differences in end-of-study means is a common goal of SMARTs. However, EDTRs may also be compared based on differences in other quantities,[4, 13] for example, *Area Under the Curve (AUC)*. AUC is defined as the total area under $\mathcal{T}^{(a_1,a_2^{NR})}(\boldsymbol{X})$ between $t_1$ and $t_T$ and can be approximated using the trapezoidal rule as

$$\sum_{\mathrm{j=1}}^{\mathrm{T-1}} \frac{1}{2}\left(\mu_j^{(a_1,a_2^{NR})}(\boldsymbol{X}) + \mu_{j+1}^{(a_1,a_2^{NR})}(\boldsymbol{X})\right)\left(t_{j+1} - t_j\right)$$

In contrast to end-of-study means, the AUC accounts for how mean trajectories of EDTRs evolve over time (see Web Figure 13). It is relevant in health domains where achieving sustained health behavior change over a longer period of time represents a clinically significant milestone (e.g., consistent therapy attendance over three months) or risk factor (e.g., consistently high alcohol use over six months).

More generally, EDTRs $\left(a_1{}', a_2^{NR}{}'\right)$ and $\left(a_1{}'', a_2^{NR}{}''\right)$ can be compared based on the contrast $\Delta_Q(\boldsymbol{X}) := \sum_{j=1}^{T} l_j \mu_j^{\left(a_1{}', a_2^{NR}{}'\right)}(\boldsymbol{X}) - \sum_{j=1}^{T} l_j \mu_j^{\left(a_1{}'', a_2^{NR}{}''\right)}(\boldsymbol{X})$ where $l_j$'s are real-valued constants. $\Delta_Q(\boldsymbol{X})$ is the difference of weighted sums of means of potential outcomes over each measurement occasion $j$, with weights given by the $l_j$'s. The choice of $l_j$'s can make $\Delta_Q(\boldsymbol{X})$ equivalent to the difference in end-of-study means (setting the $T^{th}$ weight to 1 and all others to 0), AUC (setting the first weight to $\left(\frac{t_2-t_1}{2}\right)$, the $T^{th}$ weight to $\left(\frac{t_T-t_{T-1}}{2}\right)$, and all other weights to $\left(\frac{t_{j+1}-t_{j-1}}{2}\right)$), or other estimands.

**2.3 Modeling of EDTR mean trajectories**

For illustrative purposes, we will describe modeling of EDTR mean trajectories in terms of a slightly modified, simplified version of ENGAGE. We ignore the initial two weeks before randomization and label the initial randomization as time $t_1 = 0$. We assume five subsequent measurement occasions, equally spaced at 4 weeks apart. Response status assessment and the re-randomization of nonresponders are assumed to take place at time $t_2 = 4$ weeks since initial randomization. The remaining occasions are $t_3 = 8$ weeks through $t_6 = 20$ weeks since initial randomization (see Figure 1). The actual study did not have six evenly spaced outcome measurements in this way, but they are convenient for exposition.

Our model for EDTR mean trajectories (displayed in Equation 1 below) permits each EDTR and measurement occasion to have its own parameter. Following Lu and colleagues[19], the mean trajectories of all EDTRs are constrained (a) to share the same intercept prior to randomization to first-stage options; and (b) to have identical mean trajectories until randomization to second-stage options for a pair of EDTRs that begin with the same first-stage option. Let $\mu_j^{(a_1,a_2^{NR})}(\boldsymbol{X})$ represent the mean potential outcome at time $t_j$ conditional on baseline characteristics $\boldsymbol{X}$. Here, $\boldsymbol{z}$ denotes a vector of treatment indicators in Equation 1. That is, $\boldsymbol{z} = (1, I(a_1 = +1, j = 2), \dots, I(a_1 = -1, a_2^{NR} = -1, j = T))^T$; note $\boldsymbol{z}$ is *not* a random variable. Further, $I(\cdot)$ denotes an indicator function; $m_{\boldsymbol{X}}$ and $m_{\boldsymbol{z}}$ denote the number of covariates in $\boldsymbol{X}$ and indicators in $\boldsymbol{z}$, respectively; $\boldsymbol{\beta_X} \coloneqq (\beta_{X,0}, \dots, \beta_{X,m_X})^{\boldsymbol{T}}$, $\boldsymbol{\beta_z} := (\beta_{z,0}, \dots, \beta_{z,6,T})^{\boldsymbol{T}}$ denotes a suitable link function such as the log link or logit link. Then

$$g\left(\mu_j^{(a_1,a_2^{NR})}(\boldsymbol{X})\right) := \beta_{\boldsymbol{X}}^T\boldsymbol{X} + \boldsymbol{\beta}_{\boldsymbol{z}}^T\boldsymbol{z}, \quad (1)$$

where $\boldsymbol{\beta}_{\boldsymbol{z}}^T\boldsymbol{z}$ is an underlying piecewise linear trajectory

$$\boldsymbol{\beta}_{\boldsymbol{z}}^T \boldsymbol{z} \coloneqq \beta_{z,0} + \sum_{\ell=2}^{K} I(a_1 = +1, j = \ell) \cdot \beta_{z,1,\ell}$$
$$+ \sum_{\ell=2}^{K} I(a_1 = -1, j = \ell) \cdot \beta_{z,2,\ell}$$
$$+ \sum_{\ell=K+1}^{T} I(a_1 = +1, a_2^{NR} = +1, j = \ell) \cdot \beta_{z,3,\ell}$$
$$+ \sum_{\ell=K+1}^{T} I(a_1 = +1, a_2^{NR} = -1, j = \ell) \cdot \beta_{z,4,\ell}$$
$$+ \sum_{\ell=K+1}^{T} I(a_1 = -1, a_2^{NR} = +1, j = \ell) \cdot \beta_{z,5,\ell}$$
$$+ \sum_{\ell=K+1}^{T} I(a_1 = -1, a_2^{NR} = -1, j = \ell) \cdot \beta_{z,6,\ell}$$

## 2.4 Estimation of EDTR mean trajectories and contrasts

Let $\boldsymbol{\beta} \coloneqq \begin{pmatrix} \beta_{\boldsymbol{X}} \\ \beta_{\boldsymbol{z}} \end{pmatrix}$ denote unknown parameters in the model of EDTR mean trajectories in Equation 1. To obtain a consistent estimate of $\boldsymbol{\beta}$, one can solve *inverse probability weighted and replicated* (IPWR) estimating equations, thereby fitting a single regression model to simultaneously estimate the time-varying mean outcome under all EDTRs while possibly controlling for baseline covariates.[4, 12, 19] Under usual regularity conditions (summarized in Web Appendix A), the solution $\widehat{\boldsymbol{\beta}}^{IPWR}$ to these estimating equations is a consistent estimator of $\boldsymbol{\beta}$, and $\sqrt{N}\left(\widehat{\boldsymbol{\beta}}^{IPWR} - \boldsymbol{\beta}\right)$ has an asymptotic multivariate normal distribution.

We propose a *plugin estimator* for $\Delta_Q(\boldsymbol{X})$, which we denote by $\hat{\Delta}_Q^{PI}(\boldsymbol{X})$ and obtain by simply "plugging in" $\hat{\boldsymbol{\beta}}^{IPWR}$ for $\boldsymbol{\beta}$ in the expression for $\Delta_Q(\boldsymbol{X})$. We show (Web Appendix A) that $\hat{\Delta}_Q^{PI}(\boldsymbol{X})$ is consistent for $\Delta_Q(\boldsymbol{X})$, and moreover, $\sqrt{N}\left(\hat{\Delta}_Q^{PI}(\boldsymbol{X}) - \Delta_Q(\boldsymbol{X})\right)$ has an asymptotic univariate normal distribution; this applies to various commonly used link functions $g$.

**2.5 Hypothesis testing**

We focus on sample size estimation for the case when baseline covariates $\boldsymbol{X}$ are omitted from Equation 1, and henceforth omit $(\boldsymbol{X})$ from our notation. Covariate adjustment is typically used to improve efficiency in estimating treatment effects,[20] so estimating the sample size required without covariate adjustment is generally conservative. Because omitting baseline covariates is equivalent to setting $\boldsymbol{X}$ to a vector of 1's (intercept only) and having $\boldsymbol{\beta_X} = \beta_{X,0}$ in Equation 1, the results concerning the consistency of $\hat{\boldsymbol{\beta}}^{IPWR}$ and $\hat{\Delta}_Q^{PI}$ and the asymptotic distribution of $\sqrt{N}\left(\hat{\boldsymbol{\beta}}^{IPWR} - \boldsymbol{\beta}\right)$ and $\sqrt{N}\left(\hat{\Delta}_Q^{PI} - \Delta_Q\right)$ still hold.

In this manuscript, we propose an approach to estimate sample size required for attaining a target power to reject the null hypothesis $H_0\colon \Delta_Q = 0$ against the alternative hypothesis $H_a\colon \Delta_Q \neq 0$ at type-I error rate $\alpha$. Under the null hypothesis, the test statistic $\frac{\hat{\Delta}_Q^{PI}}{\sqrt{\widehat{Var}\left(\hat{\Delta}_Q^{PI}\right)}}$ has an asymptotic $N(0,1)$ distribution (see Remark 2 in Web Appendix A).

Therefore, the test "reject $H_0$ when $\left|\frac{\hat{\Delta}_Q^{PI}}{\sqrt{\widehat{Var}\left(\hat{\Delta}_Q^{PI}\right)}}\right| > z_{1-\alpha/2}$" is asymptotically level-$\alpha$.

## 3. Considerations in developing an approach to sample size estimation

A useful sample size formula would require expressing the variance term in the denominator of the test statistic in terms of relatively interpretable quantities which can be elicited from the investigators designing the study. However, this variance term is impractical to express in a convenient closed form. Hence, we outline a customizable simulation approach to enable investigators to estimate sample size for an arbitrary number of measurement occasions $T$, specifying the data-generating model whenever possible using quantities that would be meaningful to the study designers.

### 3.1 Considerations relating to the planned SMART

It is important for the data generation step to accommodate salient features of realistic SMARTs, including multiple sequential randomizations, re-randomization based on response status, ordering of the sequential randomizations in a SMART in relation to the timing of repeated measurements, and dependency between response status and the repeated measurements.[12, 19] These features, not typical of RCTs with longitudinal data, introduce substantial complexity to the data generation step, particularly when simulating non-normal data such as overdispersed counts.

Our data-generating approach assumes that an individual is classified as a responder based on whether their observed outcome $Y_K$ at time $t_K$ is below a cutoff, above a cutoff, or whether it is between set limits. We focus on the first definition, but our approach can accommodate all three (see Table 2).

### 3.2 Considerations relating to elicitation

Study designers may find quantities at the level of an EDTR somewhat abstract and difficult to specify in advance. They may find it easier to specify quantities at the level of an *Embedded Treatment Sequence (ETS)*, a sequence of treatments actually offered to an individual. Starting at time $t_K$ (recall that $K$ denotes the specific measurement occasion immediately prior to second-stage randomization), an ETS is conditional on response status while an EDTR is marginal over response status. Hence, the mean of a longitudinal outcome at the level of an ETS is conditional on $a_1, a_2^{NR}$, and response status $R_i^{(a_1)}$ together, but the mean of a longitudinal outcome at the level of an EDTR is conditional on $(a_1, a_2^{NR})$ and marginal over response status. Quantities at the level of an ETS can seem more concrete and intuitive because they do not require averaging over response status. Hence, the design parameters (inputs to sample size estimation) of our proposed approach will be formulated at the level of an ETS, rather than an EDTR.

Second, non-normal longitudinal outcomes are often posited to come from a multivariate distribution whereby correlations, variances, and other quantities involving higher-order moments of the outcomes are functions of the outcomes' means (e.g., for overdispersed count data).[21] Investigators dealing with normal distributions typically specify the correlation structure of the longitudinal outcome of interest independently of other design parameters. However, in the case of non-normal distributions, separately specified correlations and means are not guaranteed to be jointly compatible to the posited multivariate distribution. *Copulas* – functions that link together marginal univariate cumulative distribution functions to form a multivariate cumulative distribution function (CDF) – can be utilized in a data-generative model to simulate multivariate non-normal random variables, whose marginal univariate distributions can be specified independently of the targeted correlation structure prior to data generation.

The use of copulas to simulate non-normal potential outcomes is not new. Outside the SMART setting, several authors have used a Gaussian copula to specify the joint distribution of non-normal potential outcomes (e.g., Albert and Nelson[22]). Equation 3 defines a Gaussian copula where $Y_1, \dots, Y_d$ represent components of the multivariate random variable $\boldsymbol{Y}$ of interest, $F_{Y_\ell}$'s are marginal univariate CDFs of the $Y_\ell$'s, $\Phi_d$ denotes the standard $d$-dimensional multivariate normal CDF, $\phi$ denotes the univariate standard normal CDF, and $\boldsymbol{\Lambda}_d$ denotes a $d \times d$ positive definite symmetric matrix. When the joint distribution of the multivariate random variable of interest $\boldsymbol{Y}$ is defined in terms of a Gaussian copula, $\boldsymbol{Y}$ can be viewed as a nonlinear transformation of a latent multivariate normal (MVN) random variable (say, $\boldsymbol{Z} = (Z_1, \dots, Z_d)^T$) with marginal variances 1 and Pearson correlation matrix $\boldsymbol{\Lambda}_d$. The correlations among the latent variables indirectly govern the association among the observed variables, although they are not equal. We will employ a Gaussian copula[23] in our data generation step:

$$F_{Y_1 \dots Y_d}(y_1 \dots y_d) = \Phi_d\left(\phi^{-1}\left(F_{Y_1}(y_1)\right), \quad \dots, \quad \phi^{-1}\left(F_{Y_d}(y_d)\right); \boldsymbol{\Lambda}_d\right) \tag{3}$$

## 4. Methods

We therefore combine ideas from two areas in the statistical literature, namely, the potential outcome framework and copulas, into an approach that permits simulation of non-normal data from a two-stage restricted SMART as in Figure 1.

Using copulas in data generation requires generating each component of a multivariate random variable jointly, rather than sequentially. Hence, copulas cannot be used to directly simulate observed outcomes (i.e., $Y_{i,j}$'s) in a SMART since it is impossible to determine the sequence of treatments that would be offered to an individual prior to first-stage randomization. However, it is possible to enumerate an individual's full set of potential outcomes with respect to each ETS prior to data generation by viewing each individual as belonging to one of four mutually exclusive subgroups. Analogous to the *"always survivor," "protectable," "defier," and "never survivor"* subgroups of Frangakis, et al.,[24] we introduce subgroup definitions suited to the SMART context (Section 4.2). An individual's complete set of potential outcomes can subsequently be enumerated by specifying all the potential outcomes relevant to that individual's subgroup. Hence, the data generating model requires an appropriate Gaussian copula for potential outcomes under each subgroup. The use of subgroups here is purely a data generation strategy, not an attempt to facilitate separate inferences for each subgroup.

**4.1 Notation for quantities associated with ETS**

Let $Y_{i,j}^{s}$ denote an outcome of individual $i$ that would have been observed at the $j^{th}$ measurement occasion had the individual undergone ETS $s$, and let $\mu_j^s$ denote $E\left[Y_j^s\right]$. There are four EDTRs in the SMART we consider (see Table 1). However, the number of ETSs associated with time $t_j$ is determined by when $t_j$ occurred in relation to first- and second- stage randomization. Prior to first-stage randomization, an individual would not be offered any of the treatment options, hence, we denote an ETS at $t_1$ as $(\cdot)$. After first-stage randomization but prior to second-stage randomization only the first-stage treatment would have been offered; hence, we denote an ETS at $t_j$ when $t_2 \leq t_j \leq t_K$ as $(a_1)$ and observe that there are two possible ETSs between first- and second- stage randomization. After the assessment of response status, second-stage treatment would have been offered to non-responders, whereas responders would not get re-randomized. Hence, there are six possible ETSs $(a_1, r, a_2^{NR})$ after second-stage randomization (two for responders [$r$=1; denoted by $(a_1, 1, a_2^{NR})$] and four for non-responders [$r$=0; denoted by $(a_1, 0, a_2^{NR})$]). By convention, we set $a_2^{NR} = 0$ for those individuals not re-randomized, and $a_2^{NR} = 1$ or $-1$ for those re-randomized.

The estimand $\Delta_Q$ can now be expressed in terms of quantities associated with ETSs. Let $p$ and $q$ denote the proportion of responders expected under first-stage options $a_1{}'$ and $a_1{}''$, respectively. Equation 4 shows that the desired magnitude of $\Delta_Q$ in the planned SMART is implied by eliciting $p$, $q$, and $\mu_j^S$'s (either directly or indirectly) from investigators.

$$\Delta_Q \coloneqq \sum_{j=2}^{K} l_j \left[\mu_j^{(a_1{}')}\right] - \sum_{j=2}^{K} l_j \left[\mu_j^{(a_1{}'')}\right] + \sum_{j=K+1}^{T} l_j \left[p\mu_j^{(a_1{}',1,0)} + (1-p)\mu_j^{\left(a_1{}',0,a_2^{NR'}\right)}\right] - \sum_{j=K+1}^{T} l_j \left[q\mu_j^{(a_1{}'',1,0)} + (1-q)\mu_j^{\left(a_1{}'',0,a_2^{NR''}\right)}\right] \quad (4)$$

Our notation in Equation 4 also accommodates the case when $a_1{}'$ and $a_1{}''$ are identical (i.e., the pairwise comparison of EDTRs that only differ in second-stage treatment options). In this case, $p = q$, and

$$\Delta_Q = \sum_{j=K+1}^{T} l_j (1-p) \left[\mu_j^{\left(a_1{}',0,a_2^{NR'}\right)} - \mu_j^{\left(a_1{}'',0,a_2^{NR''}\right)}\right].$$

### 4.2 Data generation

To simplify exposition, within this subsection, we suppose that only three measurement occasions are to be collected (i.e., $T$=3), but the approach is applicable to more measurement occasions. Specifically, suppose measurements are to be collected just prior to first-stage randomization (at $j = 1$, $t_1 = 0$ months), just prior to second-stage randomization (at $j = 2$, $t_2$=1 month after initial randomization), and following second-stage randomization (at $j = 3$, $t_3$=2 months after initial randomization).

***4.2.1 Enumerating an individual's complete set of potential outcomes***.

Each individual entering a two-stage restricted SMART can be thought of as belonging to one of four mutually exclusive subgroups based on whether they would respond to each first-stage option: *subgroup 1* are those who would respond to both $a_1 = +1$ and $a_1 = -1$; *subgroup 2* are those who would respond to $a_1 = +1$ but not to $a_1 = -1$; *subgroup 3* are those who would not respond to $a_1 = +1$ but would respond to $a_1 = -1$; *subgroup 4* are those who would not respond to either $a_1 = +1$ and $a_1 = -1$. It is impossible for individuals who are members of subgroup 1 to undergo the sequence $(a_1 = +1, r = 0, a_2^{NR} = +1)$, and hence, their potential outcome $Y_{i,3}^{(a_1=+1, r=0, a_2^{NR}=+1)}$ is undefined. An analogous observation could be made involving other subgroups and times $t_j$ (see Web Table 2).

### *4.2.2 Specifying the joint distribution of an individual's complete set of potential outcomes.*

Let $\boldsymbol{\theta}_i^{\mathcal{g}}$ denote a vector of potential outcomes of individual $i$ in subgroup $\mathcal{g}$. A Gaussian copula is used to specify a multivariate CDF for $\boldsymbol{\theta}_i^{\mathcal{g}}$ such that its marginal distributions (i.e., the univariate distribution of a specific potential outcome in $\boldsymbol{\theta}_i^{\mathcal{g}}$, marginal over all the other potential outcomes) adhere to a desired univariate CDF. Since the dimension of $\boldsymbol{\theta}_i^{\mathcal{g}}$ differs across subgroups, a different copula should be specified for each subgroup. Specifically, $\boldsymbol{\theta}_i^{\mathcal{g}}$ is 5-dimensional for subgroup 1 ($\mathcal{g}$=1), 6-dimensional for subgroups 2 and 3 ($\mathcal{g}$=2 and 3), 7-dimensional for subgroup 4 ($\mathcal{g}$=4).

**Specifying subgroup-specific univariate CDFs in Equation 3:** The univariate CDFs $F_{Y_\ell}$'s in Equation 3 must reflect constraints in the SMART design. Because response status is assumed to be defined in terms of a cutoff on $Y_{i,K}^{(a_1)}$ (e.g., definition (a) in Table 2), the maximum or minimum outcome at time $t_K$ for a subgroup is constrained. For example, for Subgroup 2 (i.e., under definition (a), individuals who would respond to $a_1 = +1$ but not to $a_1 = -1$), the value of $Y_{i,K}^{(+1)}$ is at most the cutoff $c$ while the value of $Y_{i,K}^{(-1)}$ is greater than $c$. The constraints on $Y_{i,K}^{(+1)}$ and $Y_{i,K}^{(-1)}$ for the remaining subgroups can be analogously specified (see Web Table 3). These constraints thus inform our choice of univariate CDFs (i.e., the $F_{Y_\ell}$'s) in the subgroup-specific copula. For example, for individuals in Subgroup 2, since $Y_{i,K}^{(+1)}$ and $Y_{i,K}^{(-1)}$ obey constraints $Y_{i,K}^{(+1)} \leq c$ and $Y_{i,K}^{(-1)} > c$, respectively, then we might choose univariate CDFs for $F_{Y_{i,K}^{(+1)}}$ and $F_{Y_{i,K}^{(-1)}}$ such that their corresponding probability mass or density function has positive support only on $(-\infty, c]$ and $(c, +\infty)$, respectively.

**Specifying subgroup-specific univariate CDF's in Equation 3 when the longitudinal outcome is overdispersed count – a case study:** In randomized trials, the negative binomial (NB) distribution has often been used to characterize overdispersed count data.[25-27] The NB distribution is given by Equation 5 where a univariate random variable $Y$ denotes the number of events during a specified time period, $\Gamma(\cdot)$ is the gamma function, and $\zeta$ is the *dispersion parameter*. When the distribution of $Y$ is given by Equation 5, its mean and variance are $\mu$ and $\mu + \zeta\mu^2$, respectively, and we write $Y \sim \text{NB}(\mu, \zeta)$.

$$f_{NB(\mu,\zeta)}(y) = \frac{\Gamma(\zeta^{-1} + y)!}{\Gamma(\zeta^{-1})!\, y\,!} \pi^{(\zeta^{-1})} (1-\pi)^y \text{ where } \pi = \frac{\zeta^{-1}}{\mu + \zeta^{-1}} \tag{5}$$

For potential outcomes corresponding to ETS $s$ at time $t_K$, we specify the univariate CDF $F_{X_\ell}$ as a truncated NB CDF defined by $F(w^*) := \Pr\{W \leq w^*\} = \sum_{i=0}^{w^*} f(w)$. Equations 6 and 7 define the function $f(w)$ if the constraints on $Y_{i,K}^s$ involve '$\leq$' or '$>$', respectively. For Equation 6 with $c = 0$, $f(w)$ reduces to a point mass at zero. Hence, the distribution corresponding to the univariate CDFs for $F_{Y_K^{(+1)}}$ and $F_{Y_K^{(-1)}}$ has positive support within $[0, c]$ and $(c, +\infty)$, respectively. For potential outcomes corresponding to ETS $s$ at any other time $t_j$, we specify the univariate CDF $F_{X_\ell}$ as a NB CDF defined by $F(w^*) := \Pr\{W \le w^*\} = \sum_{w=0}^{w^*} f_{NB\left(\mu_j^s, \zeta_j^s\right)}(w); j \neq K$. Complete specification of the univariate CDFs of the copula for each subgroup is given in Web Tables 4 and 5.

$$f(w) = \frac{f_{NB(\mu_K^s, \zeta_K^s)}(w) I(w \le c)}{\sum_{y=0}^{c} f_{NB}(\mu_K^s, \zeta_K^s)(y)} \tag{6}$$

$$f(w) = \frac{f_{NB(\mu_K^s, \zeta_K^s)}(w) I(w > c)}{1 - \sum_{y=0}^{c} f_{NB(\mu_K^s, \zeta_K^s)}(y)} \tag{7}$$

**Subgroup-specific structure for the association among components of the latent outcome:** We will specify either an order-1 autoregressive (AR1) or an exchangeable structure for the pairwise association among components of the latent outcome. Specifically, an AR1 structure specifies $Corr\left(Z_{j_1}^{s'}, Z_{j_2}^{s''}\right) = \rho^{|t_{j_1} - t_{j_2}|}$ and an exchangeable structure specifies $Corr\left(Z_{j_1}^{s'}, Z_{j_2}^{s''}\right) = \rho$ for all ETSs $s'$ and $s''$ on the same path leading to cells A-F in Figure 1 (e.g., the ETS $s' = (+1)$ at $t_2$ and the ETS $s'' = (+1,0,+1)$ at $t_3$ lie along the same path leading to cell B). We refer to $\rho$ (the correlation among the latent normal variables) as the *copula dependence parameter*. Finally, we set $Corr\left(Z_{i,j_1}^{s'}, Z_{i,j_2}^{s''}\right) = \eta$, for all ETSs $s'$ and $s''$ on different paths (e.g., the ETS $s' = (+1)$ at $t_2$ is on the path leading to cell B but the ETS $s'' = (-1,0,+1)$ at $t_3$ is on the path leading to cell E). Thus, we distinguish pairwise association between components of the latent outcome corresponding to ETS which can be feasibly observed together during the conduct of a trial (i.e., are on the same path), from ETS which cannot (are on different paths).

The complete set of potential outcomes of an individual belonging to subgroup 4 (i.e., $g = 4$) is $\boldsymbol{\theta}_i^4 = \left(Y_{i,1}^{(\cdot)}, Y_{i,2}^{(+1)}, Y_{i,2}^{(-1)}, Y_{i,3}^{(+1,0,+1)}, Y_{i,3}^{(+1,0,-1)}, Y_{i,3}^{(-1,0,+1)}, Y_{i,3}^{(-1,0,-1)}\right)^T$. Hence, the structure of $\boldsymbol{\Lambda}_7$ is given by Equation 8 when an AR1 structure is specified. The complete structure of $\boldsymbol{\Lambda}_d$ for individuals in other subgroups can be analogously written.

$$\boldsymbol{\Lambda}_7 = Corr(\boldsymbol{Z}_i^4) = Corr\begin{pmatrix} Z_{i,1}^{(\cdot)} \\ Z_{i,2}^{(+1)} \\ Z_{i,2}^{(-1)} \\ Z_{i,3}^{(+1,0,+1)} \\ Z_{i,3}^{(+1,0,-1)} \\ Z_{i,3}^{(-1,0,+1)} \\ Z_{i,3}^{(-1,0,-1)} \end{pmatrix} = \begin{pmatrix} 1 & \rho & \rho & \rho^2 & \rho^2 & \rho^2 & \rho^2 \\ \rho & 1 & \eta & \rho & \rho & \eta & \eta \\ \rho & \eta & 1 & \eta & \eta & \rho & \rho \\ \rho^2 & \rho & \eta & 1 & \eta & \eta & \eta \\ \rho^2 & \rho & \eta & \eta & 1 & \eta & \eta \\ \rho^2 & \eta & \rho & \eta & \eta & 1 & \eta \\ \rho^2 & \eta & \rho & \eta & \eta & \eta & 1 \end{pmatrix} \quad (8)$$

**Choosing the copula dependence parameter $\boldsymbol{\rho}$ after an AR1 or exchangeable structure has been specified for $\boldsymbol{\Lambda}_d$:** We may characterize pairwise association between simulated potential outcomes (i.e., $Y_{i,j}^{s}$'s) using a measure that does not directly involve any of the parameters in $\boldsymbol{\Lambda}_d$, but only quantities that investigators can interpret relatively easily. One example of an interpretable quantity is $\tau_{MAX}$, the *maximum within-person association* among longitudinal outcomes across all pairs of ETSs $s'$ and $s''$ and measurement occasions $j'$ and $j''$,

$$\tau_{MAX} := \max_{s',s'',j',j''} \left\{Corr\left(Y_{j'}^{s'}, Y_{j''}^{s''}\right)\right\}. \quad (9)$$

For sample size estimation, we are mainly interested in $\tau_{MAX}$, rather than $\mathbf{\Lambda}_d$. Similar to common practice in the Generalized Estimating Equation (GEE) literature, we conceptualize pairwise associations even among discrete outcomes in terms of a Pearson correlation.[28, 29] Hence, $Corr$ in Equation 9 is Pearson correlation, regardless of the type of longitudinal outcome (e.g., count, continuous, binary).

The pairwise association between the $Y_\ell$'s in Equation 3 is determined not only by $\mathbf{\Lambda}_d$, but also by the marginal univariate CDFs $F_{Y_\ell}$. In general, the pairwise association between the $Y_\ell$'s in Equation 3 must be estimated using Monte Carlo methods.[26] Hence, although an analytical formula relating $\mathbf{\Lambda}_d$ and $\tau_{MAX}$ is generally unavailable, we numerically estimate their relationship as a preliminary step (see Web Appendix E).

To simplify this numerical estimation, $\rho$ and $\eta$ will be shared across all four subgroups. Simulations (see Web Appendix H) suggest that, given a particular value of $\rho$, power is not sensitive to the actual value of $\eta$, so we set $\eta = \frac{\rho}{2}$ as our working assumption.

Hence, given a proposed structure such as exchangeable or AR1, only the relationship between $\rho$ and $\tau_{MAX}$ needs to be numerically estimated. Our intuition is that it is easier to elicit a maximum observed correlation $\tau_{MAX}$ and a simple approximate correlation structure, than to elicit values for many different correlation parameters at once, especially those of latent variables.

When estimating sample size, the chosen value of $\rho$ will be the specific value associated with the $\tau_{MAX}$ provided by investigators (see Web Appendix E).

***4.2.3 Simulating sequential randomizations to generate an individual's observed outcomes***

Once the copula for each subgroup is specified, the method of Madsen and Birkes[30] can be applied to simulate multivariate non-normal data (see Web Appendix D). After generating the complete set of potential outcomes for each of the $N$ individuals, observed outcomes can be chosen based on simulated sequential randomizations and the potential outcome framework's consistency assumption.[17] This assumption states that an individual's observed outcome is equal to their potential outcome under the treatment offered during the actual conduct of the trial. Thus, for the particular values $(a_1)$ and $(a_1, r, a_2^{NR})$ for individual $i$, we have that $Y_{i,1} = Y_{i,1}^{(\cdot)}$, $Y_{i,2} = Y_{i,2}^{(a_1)}$, $R_i = R_i^{(a_1)}$, $Y_{i,3} = Y_{i,3}^{(a_1,1,0)}$ if $R_i = 1$, and $Y_{i,3} = Y_{i,3}^{(a_1,0,a_2^{NR})}$ if $R_i = 0$.

***4.2.4 Specifying the number of individuals to generate per subgroup***

Let $n_{\mathcal{g}}$ denote the number of individuals that would belong to subgroup $\mathcal{g}$, let $\lceil \cdot \rceil$ denote the ceiling function, and let $p$ and $q$ denote the proportion of responders to first-stage treatment options $a_1 = +1$ and $a_1 = -1$, respectively. Under the working assumption that the number of individuals in subgroup 4 equals the minimum of the number of non-responders to either first-stage treatment, the number of individuals to generate for each subgroup can be calculated as $\lceil n_{g} \rceil$ where the $n_{g}$'s satisfy the following: $N = \sum_{g=1}^{4} n_{g}$, $p = \frac{n_1+n_2}{N}$, $q = \frac{n_1+n_3}{N}$, and

$$n_4 = \min(N(1-p), N(1-q)).$$

Our working assumption is equivalent to setting $n_4$ to its maximum possible value; the minimum is zero. Web Appendix I shows that power is not sensitive to the value of $n_4$.

### 4.3 Power calculation for a fixed sample size N

Let $\mathbb{P}_M$ denote an empirical mean across $M$ simulated datasets. For a fixed sample size $N$, we may estimate power by $\mathbb{P}_M\left[\left|\frac{\hat{\Delta}_Q^{PI}}{\sqrt{\widehat{Var}\left(\hat{\Delta}_Q^{PI}\right)}}\right| > z_{1-\alpha/2}\right]$ (see Web Appendix C).

### 4.4 Parameters elicited from investigators

Parameters elicited from investigators are:

1. Total number of individuals $N$; the criterion $\Delta_Q$; desired type-I error rate $\alpha$.
2. $c$, the cutoff used to determine response status to first-stage treatment options.
3. $\tau_{MAX}$ (see Equation 9)
4. The parameters required to specify (directly or indirectly) the means and variances of the univariate marginal distributions of the longitudinal outcomes for each subgroup $g$.

**Eliciting quantities in item 4 when the longitudinal outcome is overdispersed count – a case study:** Recall that the subgroup-specific $F_{Y_\ell}$'s in Equation 3 were specified as truncated NB or NB CDFs when the longitudinal outcome is overdispersed count. The specific parameters elicited in item 4 would then be the means and dispersion parameters (i.e., the $\mu_j^s$'s and $\zeta_j^s$'s). Investigators may specify the dispersion parameter in these CDFs directly based on prior studies, or the indirectly by specifying either the variances or the proportion of zeros in the outcome (i.e., $Var\left[Y_j^s\right]$ or $Pr\left[Y_j^s = 0\right]$, respectively). Since variances are more abstract and hence typically more challenging to elicit than proportion of zeros, we recommend eliciting the latter. The value of the dispersion parameter can be obtained by solving $Pr\left[Y_j^s = 0\right] = \left(\frac{\frac{1}{\zeta_j^s}}{\mu_j^s + \frac{1}{\zeta_j^s}}\right)^{1/\zeta_j^s}$, which follows from Equation 5. The proportion of responders to first-stage treatment $a_1$ is then implied by the elicited value of $\mu_K^{(a_1)}$ and $\zeta_K^{(a_1)}$ since $Pr\left[R^{(a_1)} = 1\right] = Pr\left[Y_K^{(a_1)} \leq c\right] = \sum_{y=0}^{c} f_{NB\left(\mu_K^{(a_1)}, \zeta_K^{(a_1)}\right)}(y)$.

## 5. Simulation Study Design

Consider a hypothetical two-stage restricted SMART (exemplified by the simplified ENGAGE SMART in Figure 1) where the longitudinal outcome of interest is overdispersed count. There are six monthly measurement occasions (i.e., $T = 6$), non-responders are re-randomized immediately after the second measurement occasion (i.e., $K = 2$), and response status is defined as $R_i^{(+1)} = I\left(Y_{i,K}^{(+1)} = 0\right)$ and $R_i^{(-1)} = I\left(Y_{i,K}^{(-1)} = 0\right)$. The pair of EDTRs (+1,+1) and (-1,+1) are compared using either the difference in end-of-study means, $\Delta_{\text{EOS}} = \mu_6^{(+1,+1)} - \mu_6^{(-1,+1)}$ , or in AUC, $\Delta_{\text{AUC}} \approx \sum_{\text{j}=1}^{6-1} \frac{1}{2}\left(\mu_j^{(a_1,a_2^{NR})} + \mu_{j+1}^{(a_1,a_2^{NR})}\right)\left(t_{j+1} - t_j\right)$, each at a desired type-I error rate of $\alpha = 0.05$. We use $p$ and $q$ to denote the proportion of responders to first-stage treatment options $a_1 = +1$ and $a_1 = -1$, respectively.

In ***Simulation Study 1***, we investigate how power changes as $\Delta_{EOS}$ and $\Delta_{\text{AUC}}$ increase. Across all scenarios, $p$ and $q$, ETS proportion of zeros, and $\rho$ were held constant as power was calculated across the grid 100, 150, 200, … 550 for total sample size $N$. Ten scenarios corresponding to increased magnitude of $\Delta_{\text{EOS}}$ and $\Delta_{AUC}$ were considered. Negative binomial distributions were used for the longitudinal outcomes. Table 3 displays the values of parameters used the study (top panel) and the dispersion parameter at each ETS and time point (bottom panel). Altogether, these values imply the following values of $\Delta_{EOS}$ and $\Delta_{\text{AUC}}$ in scenarios 1-10: $\Delta_{\text{EOS}}$ = 0.28 (scenario 1), 0.56, 0.84, 1.12, 1.4, 1.68, 1.96, 2.24, 2.52, and 2.8 (scenario 10); and $\Delta_{AUC}$ = 0.7 (scenario 1), 1.41, 2.11, 2.81, 3.52, 4.22, 4.92, 5.63, 6.33, 7.03 (scenario 10). Two simulated power curves were calculated, one where the structure of the pairwise association among latent outcomes within each subgroup (i.e., $\boldsymbol{\Lambda}_d$'s) was AR1, and another when the structure was exchangeable. It was hypothesized that power would begin very low (near alpha) for very small effect sizes and smoothly increase towards 1 for large effect sizes. Finally, we repeated the scenarios above, but increased the value of $\rho$. The magnitudes of $\Delta_{EOS}$ and $\Delta_{\mathrm{AUC}}$ remained unchanged even at increased values of $\rho$. $M = 5000$ samples were simulated to calculate power.

In ***Simulation Study 2***, we investigate whether our simulation approach was able to generate data that is consistent with the maximum within-person association ($\tau_{MAX}$) and structure of the pairwise association specified by investigators for the longitudinal outcome. We estimated $Corr\left(Y_{j'}^{s'}, Y_{j''}^{s''}\right)$ for all pairs of ETSs $s'$ and $s''$ and pairs of measurement occasions $j'$ and $j''$ by calculating their average pairwise association across $M = 5000$ simulated datasets using scenarios identical to Simulation Study 1, except that sample size was fixed to 1000.

## 6. Simulation Study Results

Power curves from Simulation Study 1 are displayed in Figure 2. They indicate that when $\rho$ was fixed to 0.2, power increases as $\Delta_{EOS}$ or $\Delta_{\mathrm{AUC}}$ increases. We observed similar trends when $\rho$ was fixed to 0.4, and 0.6. To save space, we omit some power curves in Figure 2 and display them in Web Figure 9 and Web Figure 10.

Results from ***Simulation Study 2*** show that when $\rho$ was 0.2, 0.4, 0.6, the corresponding value of $\tau_{MAX}$ was 0.15, 0.32, 0.52 in all scenarios, regardless of whether an AR1 or an exchangeable structure was utilized for the $\boldsymbol{\Lambda}_d$'s. That is, in all scenarios utilizing either type of structure for the $\boldsymbol{\Lambda}_d$'s (i.e., AR1 or exchangeable), a monotone relationship between $\rho$ and $\tau_{MAX}$ was observed.

In each scenario, the specification of an AR1 or exchangeable structure for the pairwise association among latent outcomes in each subgroup led to near-AR1 or near-exchangeable structure for the pairwise association in the longitudinal outcome. The strength of the association is bounded above by, but not identical to, the specified $\rho$. We display the estimated structure of the pairwise association in the longitudinal outcome (for the $Y_{i,j}^{S}$'s), for the case when $\rho = 0.6$ in scenario 10, in Web Figure 11 and Web Figure 12, for the AR1 and exchangeable structure, respectively. The estimated pairwise association structures for all other scenarios were similar and hence omitted.

Together, Simulation Studies 1 and 2 show that higher values of $\tau_{MAX}$ (more highly correlated outcomes) can lead to higher power for detecting differences in end-of-study means. However, these gains were observed primarily when an exchangeable structure is utilized, and not AR1 (top panels, Figure 2). On the other hand, higher values of $\tau_{MAX}$ can lead to attenuation in power for detecting differences in AUC. This attenuation was observed in both AR1 and exchangeable structures (bottom panels, Figure 2). We provide an explanation for this result below.

## 7. Discussion

### 7.1 Contribution and Further Discussion of Simulation Study Results

The current manuscript introduces a customizable Monte Carlo-based approach for estimating the sample size needed to leverage longitudinal outcome data to compare two EDTRs in terms of end-of-study outcome and AUC, and more generally in terms of $\Delta_Q$, the difference of weighted sums of means of potential outcomes over each measurement occasion. The contribution of this manuscript is twofold. First, the new Monte Carlo-based approach enables investigators to plan sample size for longitudinal data that is not continuous, while specifying design parameters that are interpretable and accommodating realistic design features of SMARTs. Second, while our sample size planning approach can be employed to a wide variety of longitudinal outcomes (and any number of measurement occasions), we demonstrated its application using a case study with a longitudinal overdispersed count outcome. This closes an important gap in sample size planning resources for longitudinal SMART studies, which are currently available primarily for linear models of continuous data[12] (Seewald et al., 2020) and, to a limited extent, to binary data.[10, 11]

Simulation studies indicate that our proposed approach possesses desirable characteristics. Specifically, simulation studies show little sensitivity of power to violations of working assumptions (see Web Appendix H and I) and anticipated increases in power as the differences in end-of-study means (i.e., $\Delta_{\mathrm{EOS}}$) or in AUC (i.e., $\Delta_{\mathrm{AUC}}$) increase. In Section 6, we observed that higher values of the maximum pairwise association $\tau_{MAX}$ can lead to attenuation in power for detecting differences in AUC.

This is because AUC is analogous to a within-cluster average, and a large pairwise association between repeated measurements represents less independent information per cluster (here, per person). In contrast, comparing end-of-study means with randomized data, under the assumption of group equivalence prior to randomization, is more similar to analysis of covariance, where a large pairwise association between repeated measurements leads to a smaller standard error. Although gains in power were observed for detecting differences in end-of-study means when an exchangeable structure was utilized, no gains in power were observed when an AR1 structure was utilized. The latter observation can be attributed to the fact that pairwise association between repeated measurements having increased time-separation are generally near-zero (e.g., $0.52^{|2-1|} = 0.52$ but $0.52^{|6-1|} = 0.04$). Although pairwise association between repeated measurements often decrease with increased time-separation among measurement occasions, within-person correlation rarely approaches zero, even if they are taken many years apart.[29] Hence, utilizing an exchangeable structure in the data-generation process may more closely mimic data collected from a broader number of SMART studies compared to an AR1 structure. Particularly when more frequent measurement occasions are anticipated, practitioners employing our proposed approach may consider using an exchangeable structure but calculate power across different values for $\tau_{MAX}$ as a kind of sensitivity analysis.

In the simulation studies, we investigated the properties of our approach with overdispersed count data. We conjecture that as long as the marginal univariate CDF's in the Gaussian copula (i.e., $F_{Y_\ell}$'s in Equation 3) belong to the class of dispersion models (DM's),[23] the properties of our proposed approach will generalize; this class includes the Gaussian, gamma, and inverse gamma (continuous) distributions and the binomial, Poisson, negative binomial (discrete) distributions as important special cases.

### 7.2 Limitations and Directions for Future Research

There are several limitations to the proposed work. First, in simulation studies considering the case when the longitudinal outcome of interest is overdispersed count and with finite sample sizes ranging from 100 to 550 (see Web Appendix F), we found the empirical type-I error rate for the proposed test to be either nominal or slightly above nominal (i.e., about 0.05 to 0.07) when comparing EDTRs based on differences in end-of-study means. However, when comparing EDTRs based on differences in AUC, empirical type-I error rate was sometimes slightly above nominal (0.07 – 0.09). This occurred in scenarios where overdispersion was more extreme and sample size was 200 or less. The likely explanation is that the estimate $\widehat{Var}(\hat{\Delta}_Q^{PI})$ of the sampling variance of the estimand of interest, obtained from Taylor series arguments, can be biased in such extreme scenarios (see Web Appendix G). A bootstrap-based approach to obtaining $\widehat{Var}(\hat{\Delta}_Q^{PI})$ might improve accuracy[31] in future work.

Second, we followed common practice in the GEE literature in expressing pairwise association among repeated measurements as a Pearson correlation, even if the measurements were discrete.[28] However, pairwise association might not be fully captured by Pearson correlation when repeated measurements are discrete. Extending our approach to accommodate a broader variety of ways such pairwise associations can be conceptualized (e.g., polychoric correlation[32]) is an area for future work.

Third, outside the SMART literature, confidence intervals[33] (e.g., in GLIMMPSE software[34]) are sometimes calculated for power (given a fixed sample size N) to capture the uncertainty in the elicited parameters. As an area for future research, our proposed approach may be extended to provide confidence intervals for estimates of power.

Finally, we did not discuss missing data or attrition in this paper, but it should be considered when planning a study, because it reduces power or increases initial sample size requirements. In Web Appendix J, we explain how to add missingness and multiple imputation to the data generation step.

**7.3 Conclusion**

This manuscript closes an important gap in sample size planning resources for SMART studies. The flexible Monte Carlo-based approach developed here enables researchers to plan sample sizes for comparing EDTRs based on a wide variety of longitudinal outcomes, most notably based on overdispersed count outcomes. This approach focuses on the ETS when eliciting parameters from investigators, even though the estimand of interest (i.e., $\Delta_Q$) focuses on the EDTR. In the context of sample size estimation for SMARTs, some authors[13] have also required elicitation of parameters at the ETS level. However, to our knowledge, none have made the explicit connection between eliciting parameters at the ETS level to interpretability and meaningfulness of parameters while accommodating realistic features of SMARTs, as in this manuscript. This approach can be helpful when developing sample size estimation methods for comparing mean trajectories of EDTRs.

**Figure 1:** The overall ENGAGE study. Shaded section represents the ENGAGE SMART. Solid black circles represent randomization to treatment options. $K$ denotes the specific measurement occasion immediately prior to randomization to second-stage treatment options; $T$ denotes the total number of measurement occasions. In the simplified ENGAGE SMART (solid black arrow), first-stage randomization occurs 2 weeks later than the actual ENGAGE SMART (represented by the solid black arrow; see also assumed versus actual timeline for $t_j$ displayed)

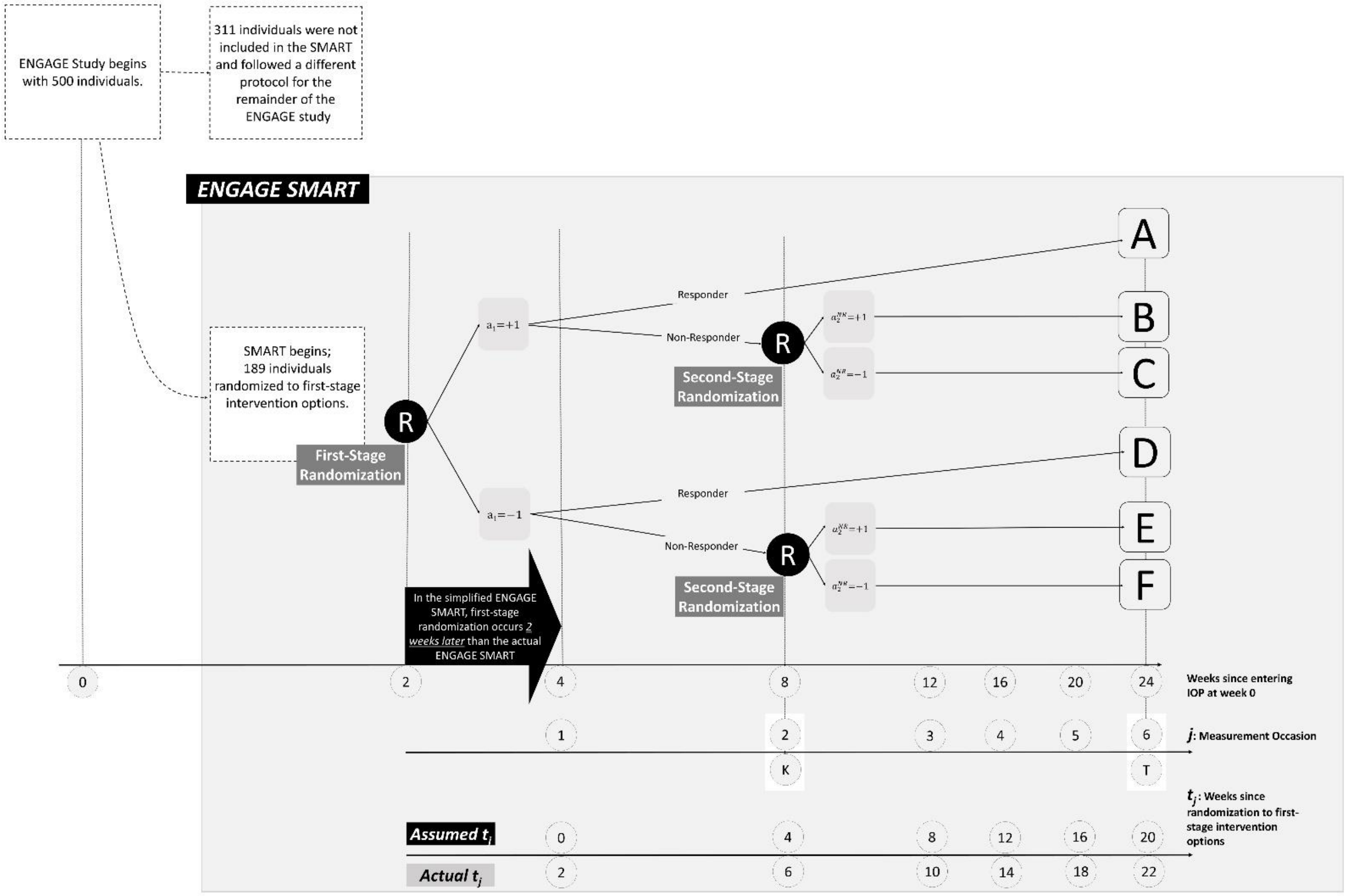

**Figure 2:** Results of Simulation Study 1 when an AR1 and exchangeable structure is utilized in the specification of the latent MVN-distributed variable are displayed on the left and right panels, respectively. For space considerations, only select power curves for difference in AUC are displayed.

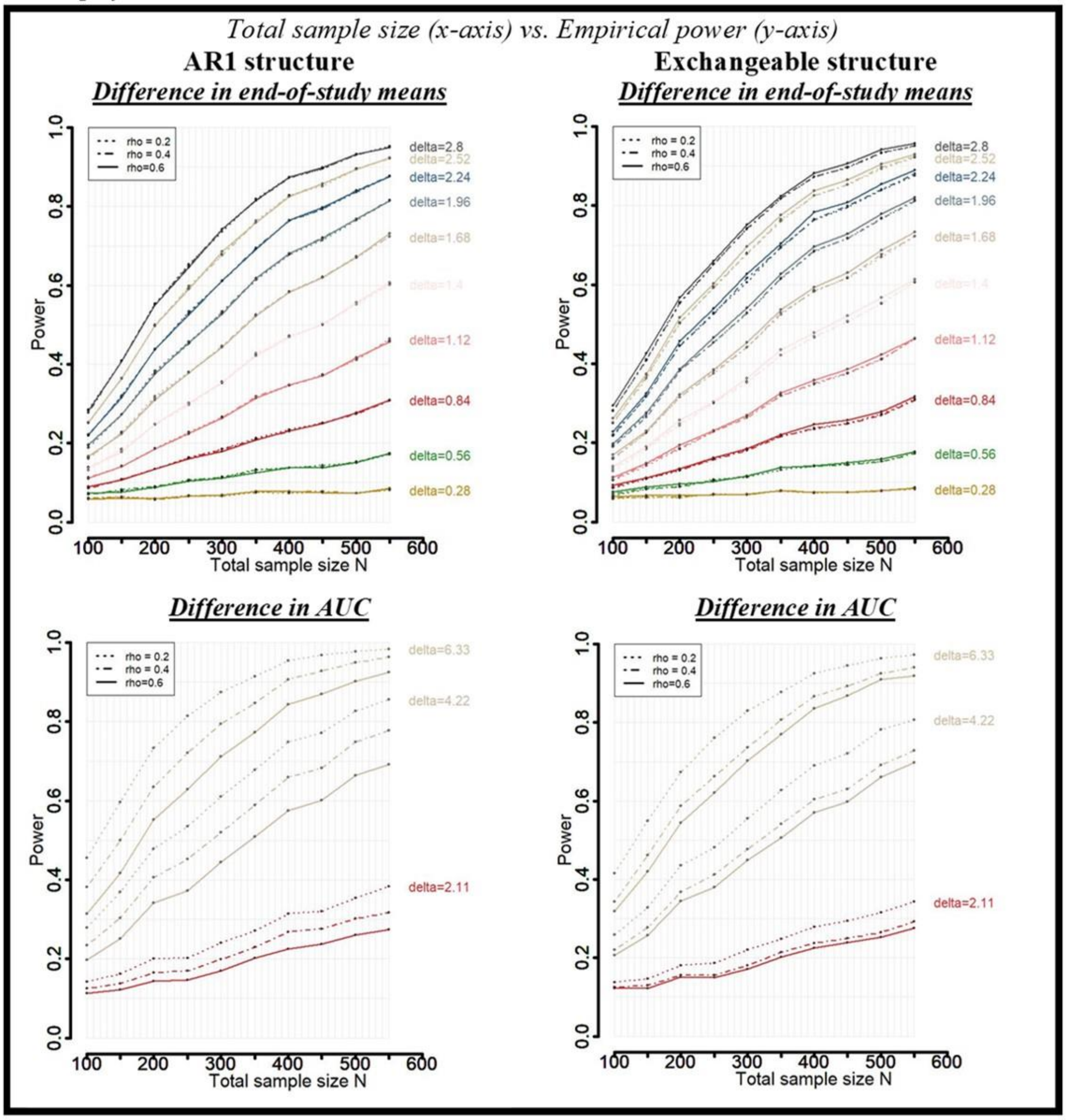

**Table 1:** EDTRs in the ENGAGE SMART

| | $(a_1, a_2^{NR})$ | Example EDTRs from the ENGAGE SMART |
|---|---|---|
| (1) | $(+1, +1)$ | ***‘Choice-Throughout’ DTR:***<br>First offer the individual MI-Choice (coded as $a_1 = +1$) at week 2.<br><br>If the individual is regarded as a *responder* at week 8 (i.e., $R_i$=1), then have no further contact by study team (NFC); else (i.e., $R_i$=0) offer MI-Choice (coded as $a_2^{NR} = +1$) again at week 8. |
| (2) | $(+1, -1)$ | ***‘Choice-Initially’ DTR:***<br>First offer the individual MI-Choice (coded as $a_1 = +1$) at week 2.<br><br>If the individual is regarded as a responder at week 8, then have NFC; else have NFC (coded as $a_2^{NR} = -1$) at week 8. |
| (3) | $(-1, +1)$ | ***‘Delayed-Choice’ DTR:***<br>First offer the individual MI-IOP (coded as $a_1 = -1$) at week 2.<br><br>If the individual is regarded as a responder at week 8, then have NFC; else offer MI-Choice (coded as $a_2^{NR} = +1$) at week 8. |
| (4) | $(-1, -1)$ | ***‘No-Choice’ DTR:***<br>First offer the individual MI-IOP (coded as $a_1 = -1$) at week 2.<br><br>If the individual is regarded as a responder at week 8, then have NFC; else have NFC (coded as $a_2^{NR} = -1$) at week 8. |

**Table 2:** Examples of how response status can be defined in a SMART with a longitudinal outcome. Recall that $K$ denote the specific measurement occasion immediately prior to randomization to second-stage treatment options.

| | Definition of response status | Examples when $Y_{i,K}^{(a_1)}$ is count |
|---|---|---|
| (a) | $R_i^{(a_1)} = I\left(Y_{i,K}^{(a_1)} \leq c\right)$ | An individual is regarded as a responder at $t_K$ if the number of alcohol-use days they had in the past-month (i.e., $Y_{i,K}^{(a_1)}$ in this example) does not exceed the cutoff $c$. If the cutoff $c$ is selected by study designers to be equal to 0, then an individual is regarded as a responder at $t_K$ if they abstained from any alcohol-use in the month prior to $t_K$. |
| (b) | $R_i^{(a_1)} = I\left(Y_{i,K}^{(a_1)} > c\right)$ | An individual is regarded as a responder if at $t_K$, the number of therapy sessions they attended in the past-month (i.e., $Y_{i,K}^{(a_1)}$ in this example) exceeds the cutoff $c$. If the cutoff $c$ is selected by study designers to be equal to 0, then an individual is regarded as a responder at $t_K$ if they attended at least one therapy session in the month prior to $t_K$. |
| (c) | $R_i^{(a_1)} = I\left(c_1 \leq Y_{i,K}^{(a_1)} \leq c_2\right)$ | An individual is regarded as a responder at $t_K$ if the number of times they adhered to a clinician-prescribed regimen of medication in the past-month (i.e., $Y_{i,K}^{(a_1)}$in this example) range between $c_1$ and $c_2$. That is, going beyond the cutoff $c_2$ signifies that the individual deviated from the prescribed regimen by taking *more* than what was prescribed. On the other hand, going below the cutoff $c_1$ signifies that the individual deviated from the prescribed regimen by taking *less* than what was prescribed. |

**Table 3:** Parameter values in simulation studies. Below, $\pi_j^s$ denotes $Pr\left[Y_j^s = 0\right]$.

***Fixed across all scenarios:***

***Total Sample Size***

N=100, 150, 200, …, 550

***Copula Dependence Parameter***

$\rho = 0.2, 0.4, 0.6$

***Proportion of Responders***

$p = 0.40$ and $q = 0.40$

***ETS Proportion of Zeros***

$\pi_1^{(\cdot)} = 0.40, \pi_2^{(+1)} = 0.40, \pi_2^{(-1)} = 0.40,$

$\pi_3^{(+1,0,0)} = \pi_3^{(+1,0,+1)} = \pi_3^{(+1,0,-1)} = 0.40, \pi_4^{(+1,0,0)} = \pi_4^{(+1,0,+1)} = \pi_4^{(+1,0,-1)} = 0.40,$

$\pi_5^{(+1,0,0)} = \pi_5^{(+1,0,+1)} = \pi_5^{(+1,0,-1)} = 0.40, \pi_6^{(+1,0,0)} = \pi_6^{(+1,0,+1)} = \pi_6^{(+1,0,-1)} = 0.40,$

$\pi_3^{(-1,0,0)} = \pi_3^{(-1,0,+1)} = \pi_3^{(-1,0,-1)} = 0.40, \pi_4^{(-1,0,0)} = \pi_4^{(-1,0,+1)} = \pi_4^{(-1,0,-1)} = 0.40,$

$\pi_5^{(-1,0,0)} = \pi_5^{(-1,0,+1)} = \pi_5^{(-1,0,-1)} = 0.40, \pi_6^{(-1,0,0)} = \pi_6^{(-1,0,+1)} = \pi_6^{(-1,0,-1)} = 0.40$

***Varied across scenarios:***

***'Base Case'***

***ETS Means***

For all s: $\mu_1^s = 2.5, \mu_2^s = 4.8, \mu_3^s = 2.6, \mu_4^s = 2.7, \mu_5^s = 2.75, \mu_6^s = 2.8$

***Scenarios 1-10***

***ETS Means***

For $j = 1,2$*: $\mu_j^s$ takes on the following values*

For all s: $\mu_1^s = 2.5, \mu_2^s = 4.8$

For $j = 3,4,5$: *$\mu_j^s$ takes on the following values*

The value of $\mu_j^{(+1,0,0)}$, $\mu_j^{(+1,0,+1)}$, and $\mu_j^{(+1,0,-1)}$ was increased by about 7%, 14%, 28%, …, 70% respectively, from the values specified in the 'Base Case' above, but the value of $\mu_j^{(-1,0,0)}$, $\mu_j^{(-1,0,+1)}$, and $\mu_j^{(-1,0,-1)}$ was retained from the values specified in the 'Base Case' above.

For $j = 6$: *$\mu_j^s$ takes on the following values*

The value of $\mu_j^{(+1,0,0)}$, $\mu_j^{(+1,0,+1)}$, and $\mu_j^{(+1,0,-1)}$ was increased by about 10%, 20%, 30%, …, 100% respectively, from the values specified in the 'Base Case' above, but the value of $\mu_j^{(-1,0,0)}$, $\mu_j^{(-1,0,+1)}$, and $\mu_j^{(-1,0,-1)}$ was retained from the values specified in the 'Base Case' above.

***Value of dispersion parameter $\zeta_j^s$ implied by choice of values for $\mu_j^s$ and $\pi_j^s$:***

For $j = 1,2$*: $\zeta_j^s$ takes on the following values throughout Scenarios 1-10*

For all s: $\zeta_1^s = 1.92, \zeta_2^s = 2.98$

For $j = 3,4,5,6$: *$\zeta_j^s$ takes on the following value throughout Scenarios 1-10*

$\zeta_3^{(-1,0,0)} = \zeta_3^{(-1,0,+1)} = \zeta_3^{(-1,0,-1)} = 1.98, \zeta_4^{(-1,0,0)} = \zeta_4^{(-1,0,+1)} = \zeta_4^{(-1,0,-1)} = 2.05,$

$\zeta_5^{(-1,0,0)} = \zeta_5^{(-1,0,+1)} = \zeta_5^{(-1,0,-1)} = 2.08, \zeta_5^{(-1,0,0)} = \zeta_6^{(-1,0,+1)} = \zeta_6^{(-1,0,-1)} = 2.11$

For $j = 3,4,5,6$: *$\zeta_{t_j}^s$ takes on the following values in Scenario 1*

$\zeta_3^{(+1,0,0)} = \zeta_3^{(+1,0,+1)} = \zeta_3^{(+1,0,-1)} = 2.10,$

$\zeta_4^{(+1,0,0)} = \zeta_4^{(+1,0,+1)} = \zeta_4^{(+1,0,-1)} = 2.16,$

$\zeta_5^{(+1,0,0)} = \zeta_5^{(+1,0,+1)} = \zeta_5^{(+1,0,-1)} = 2.19,$

$\zeta_6^{(+1,0,0)} = \zeta_6^{(+1,0,+1)} = \zeta_6^{(+1,0,-1)} = 2.27,$

For $j = 3$: Throughout Scenarios 2-10, $\zeta_j^{(+1,0,0)} = \zeta_j^{(+1,0,+1)} = \zeta_j^{(+1,0,-1)}$. The value of $\zeta_j^{(+1,0,0)}, \zeta_j^{(+1,0,+1)}$ and $\zeta_j^{(+1,0,-1)}$ was increased by about 5%, 10%, 14%, …, 36% respectively from the values in Scenario 1.

For $j = 4$: Throughout Scenarios 2-10, $\zeta_j^{(+1,0,0)} = \zeta_j^{(+1,0,+1)} = \zeta_j^{(+1,0,-1)}$. The value of $\zeta_j^{(+1,0,0)}, \zeta_j^{(+1,0,+1)}$ and $\zeta_j^{(+1,0,-1)}$ was increased by about 5%, 9%, 14%, …, 34% respectively from the values in Scenario 1.

For $j = 5$: Throughout Scenarios 2-10, $\zeta_j^{(+1,0,0)} = \zeta_j^{(+1,0,+1)} = \zeta_j^{(+1,0,-1)}$. The value of $\zeta_j^{(+1,0,0)}, \zeta_j^{(+1,0,+1)}$ and $\zeta_j^{(+1,0,-1)}$ was increased by about 5%, 9%, 13%, …, 34% respectively from the values in Scenario 1.

For $j = 6$: Throughout Scenarios 2-10, $\zeta_j^{(+1,0,0)} = \zeta_j^{(+1,0,+1)} = \zeta_j^{(+1,0,-1)}$. The value of $\zeta_j^{(+1,0,0)}, \zeta_j^{(+1,0,+1)}$ and $\zeta_j^{(+1,0,-1)}$ was increased by about 6%, 12%, 17%, …, 42% respectively from the values in Scenario 1.

**Sample size estimation for comparing dynamic treatment regimens in a SMART: a Monte Carlo-based approach and case study with longitudinal overdispersed count outcomes**

**Supplementary Material**

## Web Appendix A. Asymptotic Distribution of Estimator

In this appendix we use the Delta Method (following van der Vaart[1]) to derive the asymptotic distribution of the estimated contrast of interest $\widehat{\Delta}_{Q}^{PI}(\boldsymbol{X})$, based on the established asymptotic distribution of the inverse-probability weighted and replicated regression estimates $\widehat{\boldsymbol{\beta}}^{IPWR}$ of the model coefficients $\boldsymbol{\beta} \coloneqq \begin{pmatrix} \boldsymbol{\beta}_{\boldsymbol{X}} \\ \boldsymbol{\beta}_{\boldsymbol{z}} \end{pmatrix}$.

We use the results of Lu et al.[2] and Seewald et al.[3] for the asymptotic distribution of $\widehat{\boldsymbol{\beta}}^{IPWR}$. Let $m_{\boldsymbol{X}}$ be the number of coefficients in the longitudinal model for the effects of the baseline covariates $\boldsymbol{X}$ and let $m_{\boldsymbol{z}}$ be the number of coefficients in the model for the treatment and treatment-by-time effects, so that $m_{\boldsymbol{X}} + m_{\boldsymbol{z}}$ is the total number of regression parameters. Let $\left(a_1{'}, a_2^{NR'}\right)$ and $\left(a_1{''}, a_2^{NR''}\right)$ be a pair of EDTRs, let $\boldsymbol{L} = (l_1, \dots, l_T)$ be real valued constants, and $\boldsymbol{b} \coloneqq \begin{pmatrix} \boldsymbol{b}_{\boldsymbol{X}} \\ \boldsymbol{b}_{\boldsymbol{z}} \end{pmatrix}$ be a column vector in $\mathbb{R}^{m_X + m_z}$. As in the main manuscript, we use $g$ to denote a link function. We define the function $\varphi\colon \mathbb{R}^{m_X+m_z} \longmapsto \mathbb{R}$ as

$$\varphi(\boldsymbol{b}) \coloneqq \sum_{j=1}^{T} l_j g^{-1}\left(h_j^{\left(a_1{'}, a_2^{NR'}\right)}(\boldsymbol{b})\right) - \sum_{j=1}^{T} l_j\, g^{-1}\left(h_j^{\left(a_1{''}, a_2^{NR''}\right)}(\boldsymbol{b})\right), \tag{S1}$$

where $\boldsymbol{b}_{\boldsymbol{z}} \coloneqq (b_{z,0}, \dots, b_{z,6,T})^{\boldsymbol{T}}$, and $h_j^{(a_1, a_2^{NR})}(\boldsymbol{b}) \coloneqq \boldsymbol{b}_{\boldsymbol{X}}^{\boldsymbol{T}}\boldsymbol{X} + \boldsymbol{b}_{\boldsymbol{z}}^{\boldsymbol{T}}\boldsymbol{z}$ is the linear estimator for the fitted value of the outcome at time $t_j$ under DTR $(a_1, a_2^{NR})$. Following the structure given in the main paper, $\boldsymbol{b}_{\boldsymbol{z}}^{\boldsymbol{T}}\boldsymbol{z}$ is defined as follows:

$$\boldsymbol{b}_{\boldsymbol{z}}^{\boldsymbol{T}}\boldsymbol{z} \coloneqq b_{z,0} + \sum_{\ell=2}^{K} I(a_1 = +1, j = \ell) \cdot b_{z,1,\ell} + \sum_{\ell=2}^{K} I(a_1 = -1, j = \ell) \cdot b_{z,2,\ell}$$

$$+ \sum_{\ell=K+1}^{T} I(a_1 = +1, a_2^{NR} = +1, j = \ell) \cdot b_{z,3,\ell}$$

$$+ \sum_{\ell=K+1}^{T} I(a_1 = +1, a_2^{NR} = -1, j = \ell) \cdot b_{z,4,\ell} \quad \text{(S2)}$$

$$+ \sum_{\ell=K+1}^{T} I(a_1 = -1, a_2^{NR} = +1, j = \ell) \cdot b_{z,5,\ell}$$

$$+ \sum_{\ell=K+1}^{T} I(a_1 = -1, a_2^{NR} = -1, j = \ell) \cdot b_{z,6,\ell}$$

Having established notation, we state the following theorem.

**Theorem A1:** Let $\boldsymbol{\Sigma}_{\widehat{\boldsymbol{\beta}}^{IPWR}}$ denote the asymptotic variance of $\sqrt{N}\left(\widehat{\boldsymbol{\beta}}^{IPWR} - \boldsymbol{\beta}\right)$. If conditions **C1** and **C2** (stated below) hold, then $\widehat{\Delta}_Q^{PI}(\boldsymbol{X})$ is consistent for $\Delta_Q(\boldsymbol{X})$, and furthermore $\sqrt{N}\left(\widehat{\Delta}_Q^{PI}(\boldsymbol{X}) - \Delta_Q(\boldsymbol{X})\right)$ has the asymptotic univariate normal distribution $N\left(0, \Sigma_{\widehat{\Delta}_Q^{PI}}\right)$ where $\Sigma_{\widehat{\Delta}_Q^{PI}} \coloneqq \boldsymbol{\psi}_{\boldsymbol{\beta}} \boldsymbol{\Sigma}_{\widehat{\boldsymbol{\beta}}^{IPWR}} \boldsymbol{\psi}_{\boldsymbol{\beta}}^{T}$, and $\boldsymbol{\psi}_{\boldsymbol{\beta}}$ is a row vector denoting the gradient of φ (with respect to $\boldsymbol{b}$) evaluated at $\boldsymbol{\beta}$.

**(C1)** All partial derivatives (with respect to $\boldsymbol{b}$) of φ exist in a neighborhood of $\boldsymbol{\beta}$ and are continuous at $\boldsymbol{\beta}$.

**(C2)** The regularity conditions for the consistency and asymptotic normality of the model coefficient estimators hold. These regularity conditions are more rigorously discussed by Murphy et al.,[4] Nahum-Shani[5] et al., Lu et al.[2], and Seewald et al.[3] but we briefly sketch them out below:

i. The model for the treatment effects on the mean (but not necessarily the variance function) is correctly specified. This is satisfied if the model is saturated and there are no baseline covariates.

ii. The randomization is performed correctly so that each participant has, in advance of the study, a nonzero probability of being consistent with each EDTR.

iii. The weighted and replicated estimating functions being set to zero to estimate the $\boldsymbol{\beta}$ coefficients have a finite and invertible expected cross-product (its structure need not be known).

**Proof**.

Under C2, $\sqrt{N}\left(\widehat{\boldsymbol{\beta}}^{IPWR}-\boldsymbol{\beta}\right)$is asymptotically multivariate normal with covariance matrix $\Sigma_{\widehat{\boldsymbol{\beta}}^{IPWR}}$. Then by C1 and the Delta Method (see, e.g., van der Vaart, Theorem 3.1, p. 26), the asymptotic distribution of $\sqrt{N}\left(\varphi\left(\widehat{\boldsymbol{\beta}}^{IPWR}\right)-\varphi(\boldsymbol{\beta})\right)$ can be derived. Specifically, $\sqrt{N}\left(\varphi\left(\widehat{\boldsymbol{\beta}}^{IPWR}\right)-\varphi(\boldsymbol{\beta})\right)$ has the asymptotic univariate normal distribution $N\left(0, \boldsymbol{\psi}_{\boldsymbol{\beta}}\boldsymbol{\Sigma}_{\widehat{\boldsymbol{\beta}}^{IPWR}}\boldsymbol{\psi}_{\boldsymbol{\beta}}^{T}\right)$.

■

**Remarks**.

1. The asymptotic variance of the contrast of interest (i.e., $\widehat{\Delta}_Q^{PI}(\boldsymbol{X})$) is $\frac{1}{N}\Sigma_{\widehat{\Delta}_Q^{PI}(\boldsymbol{X})} = \frac{1}{N}\boldsymbol{\psi}_{\boldsymbol{\beta}}\boldsymbol{\Sigma}_{\widehat{\boldsymbol{\beta}}^{IPWR}}\boldsymbol{\psi}_{\boldsymbol{\beta}}^{T}$. We note that this is a scalar even though $\boldsymbol{\Sigma}_{\widehat{\boldsymbol{\beta}}^{IPWR}}$ is a larger matrix. We propose the plug-in variance estimator $\widehat{Var}\left(\widehat{\Delta}_Q^{PI}(\boldsymbol{X})\right) = \frac{1}{N}\widehat{\Sigma}_{\widehat{\Delta}_Q^{PI}} = \frac{1}{N}\boldsymbol{\psi}_{\widehat{\boldsymbol{\beta}}^{IPWR}}\widehat{\boldsymbol{\Sigma}}_{\widehat{\boldsymbol{\beta}}^{IPWR}}\boldsymbol{\psi}_{\widehat{\boldsymbol{\beta}}^{IPWR}}^{T}$ as an estimate of the true asymptotic variance $\frac{1}{N}\boldsymbol{\psi}_{\boldsymbol{\beta}}\boldsymbol{\Sigma}_{\widehat{\boldsymbol{\beta}}^{IPWR}}\boldsymbol{\psi}_{\boldsymbol{\beta}}^{T}$. In Web Appendix G, we investigate the performance of our proposed plug-in variance estimator $\widehat{Var}\left(\widehat{\Delta}_Q^{PI}(\boldsymbol{X})\right)$ when the longitudinal outcome of interest is overdispersed count.

2. It also follows from Theorem A1 that under the null hypothesis, $\frac{\hat{\Delta}_Q^{PI}}{\sqrt{\widehat{Var}\left(\hat{\Delta}_Q^{PI}\right)}}$ has an asymptotic $N(0,1)$ distribution. This allows for approximate significance testing.

Theorem A1 applies to a variety of link functions used in practice. However, the specific form of $\boldsymbol{\psi}_{\boldsymbol{\beta}}$ depends on the selected link function. In Web Appendix B below, we show the specific form of $\boldsymbol{\psi}_{\boldsymbol{\beta}}$ when a log-link function is utilized in Equation S1.

## Web Appendix B. Calculating the gradient function

We now derive the specific form of $\boldsymbol{\psi}_{\boldsymbol{\beta}}$ when the link function $g$ is the log-link as in Equation 1 of the main manuscript. First, notice that the quantity $\varphi(\boldsymbol{b})$ defined by Equations S1-S2 can be re-expressed as:

$$\varphi(\boldsymbol{b}) = \boldsymbol{L}g^{-1}\left[\boldsymbol{C}^{\left(a_1{}',a_2^{NR'}\right)}\boldsymbol{b}\right] - \boldsymbol{L}g^{-1}\left[\boldsymbol{C}^{\left(a_1{}'',a_2^{NR''}\right)}\boldsymbol{b}\right] \qquad \text{(S3)}$$

where $g^{-1}[\cdot]$ denotes element-wise evaluation of $g^{-1}$, and $\mathbf{L}$ and $\boldsymbol{C}^{(a_1,a_2^{NR})}$ are the matrices defined below:

- $\boldsymbol{L} = (l_1, \dots, l_T)$ is a $1 \times T$ matrix whose elements are the real-valued constants $l_j$'s.
- $\boldsymbol{C}^{(a_1,a_2^{NR})}$ denotes a $T \times (m_{\boldsymbol{X}} + m_{\boldsymbol{Z}})$ matrix of 1's and 0's; the specific form of $\boldsymbol{C}^{(a_1,a_2^{NR})}$ is chosen so that the term $g^{-1}\left[\boldsymbol{C}^{(a_1,a_2^{NR})}\,\boldsymbol{b}\right]$ results in a diagonal matrix having $g^{-1}\left(h_j^{(a_1,a_2^{NR})}(\boldsymbol{b})\right)$ as its $(j,j)^{th}$ element.

Specifically, when $g$ is the log-link, $g^{-1}\left[\boldsymbol{C}^{(a_1,a_2^{NR})}\,\boldsymbol{b}\right] = exp\left[\boldsymbol{C}^{(a_1,a_2^{NR})}\,\boldsymbol{b}\right]$ where $exp$ denotes elementwise exponentiation, and $\varphi(\boldsymbol{b}) = \boldsymbol{L}exp\left[\boldsymbol{C}^{\left(a_1{}',a_2^{NR'}\right)}\boldsymbol{b}\right] - \boldsymbol{L}exp\left[\boldsymbol{C}^{\left(a_1{}'',a_2^{NR''}\right)}\boldsymbol{b}\right]$.

Therefore, when $g$ is the log-link, the gradient of the quantity $\varphi(\boldsymbol{b})$ with respect to $\boldsymbol{b}$ can be expressed as:

$$\frac{\partial\varphi(\boldsymbol{b})}{\partial\boldsymbol{b}} \coloneqq \boldsymbol{D}\boldsymbol{U}_{\boldsymbol{b}}\boldsymbol{C} \tag{S4}$$

where $\boldsymbol{D}$, $\boldsymbol{U}_{\boldsymbol{b}}$, and $\boldsymbol{C}$ are the matrices defined below:

- $\boldsymbol{D} := (\boldsymbol{L} \quad -\boldsymbol{L})$
- $\boldsymbol{U}_{\boldsymbol{b}} := \begin{pmatrix} \boldsymbol{U}_{\boldsymbol{b}}^{\left(a_1{}', a_2^{NR}{}'\right)} & \boldsymbol{0}_{T\times T} \\ \boldsymbol{0}_{T\times T} & \boldsymbol{U}_{\boldsymbol{b}}^{\left(a_1{}'', a_2^{NR}{}''\right)} \end{pmatrix}$ where we use $\boldsymbol{U}_{\boldsymbol{b}}^{(a_1, a_2^{NR})}$ to denote elementwise $exp\left[\boldsymbol{C}^{(a_1, a_2^{NR})}\, \boldsymbol{b}\right]$.
- $\mathbf{C} := \begin{pmatrix} \boldsymbol{C}^{\left(a_1{}', a_2^{NR}{}'\right)} \\ \boldsymbol{C}^{\left(a_1{}'', a_2^{NR}{}''\right)} \end{pmatrix}$

Above, $\boldsymbol{D}$ is a $1 \times 2T$ matrix, $\boldsymbol{U}_{\boldsymbol{b}}$ is a $2T \times 2T$ matrix, $\boldsymbol{C}$ is a $2T \times (m_{\boldsymbol{X}} + m_{\boldsymbol{z}})$ matrix. Hence, $\boldsymbol{D}\boldsymbol{U}_{\boldsymbol{b}}\boldsymbol{C}$ is a $1 \times (m_{\boldsymbol{X}} + m_{\boldsymbol{z}})$ matrix.

When $g$ is the log-link, $\boldsymbol{\psi}_{\boldsymbol{\beta}}$ is therefore given by $\boldsymbol{\psi}_{\boldsymbol{\beta}} = \boldsymbol{D}\boldsymbol{U}_{\boldsymbol{\beta}}\boldsymbol{C}$**.** It is still necessary to calculate $\boldsymbol{D}$ and $\boldsymbol{C}$.

- Expressing $\boldsymbol{D}$: when $\Delta_Q$ represents the difference in end-of-study means, $\boldsymbol{D} = (\boldsymbol{L} \quad -\boldsymbol{L})$ where $\boldsymbol{L} = (\boldsymbol{0}_{1\times 5} \quad 1)$. Alternatively, when $\Delta_Q$ represents the difference in AUC (see Web Figure 13), $\boldsymbol{D} = (\boldsymbol{L} \quad -\boldsymbol{L})$ where

$$\boldsymbol{L} = \left(\frac{t_2 - t_1}{2}, \quad \frac{t_3 - t_1}{2}, \quad \frac{t_4 - t_2}{2}, \quad \frac{t_5 - t_3}{2}, \quad \frac{t_6 - t_4}{2}, \quad \frac{t_6 - t_5}{2}\right).$$

- Expressing $\boldsymbol{U_\beta}$: The EDTR means implied by Equation 1 of the main manuscript are given by Web Table 1. Hence, $\boldsymbol{U}_{\boldsymbol{\beta}}^{(+1,+1)}$ is given by Equation S5; the matrices $\boldsymbol{U}_{\boldsymbol{\beta}}^{(+1,+1)}$, $\boldsymbol{U}_{\boldsymbol{\beta}}^{(+1,-1)}$, $\boldsymbol{U}_{\boldsymbol{\beta}}^{(-1,+1)}$, $\boldsymbol{U}_{\boldsymbol{\beta}}^{(-1,-1)}$ are defined analogously.

$$\boldsymbol{U}_{\boldsymbol{\beta}}^{(+1,+1)} = \begin{pmatrix} e^{\beta_{z,0}} & 0 & 0 & 0 & 0 & 0 \\ 0 & e^{\beta_{z,0}+\beta_{z,1,2}} & 0 & 0 & 0 & 0 \\ 0 & 0 & e^{\beta_{z,0}+\beta_{z,3,3}} & 0 & 0 & 0 \\ 0 & 0 & 0 & e^{\beta_{z,0}+\beta_{z,3,4}} & 0 & 0 \\ 0 & 0 & 0 & 0 & e^{\beta_{z,0}+\beta_{z,3,5}} & 0 \\ 0 & 0 & 0 & 0 & 0 & e^{\beta_{z,0}+\beta_{z,3,6}} \end{pmatrix} \tag{S5}$$

- Expressing $\boldsymbol{C}$: The matrices $\boldsymbol{C}^{(+1,+1)}$, $\boldsymbol{C}^{(+1,-1)}$, $\boldsymbol{C}^{(-1,+1)}$, $\boldsymbol{C}^{(-1,-1)}$ are defined as:

$$\boldsymbol{C}^{(+1,+1)} := (\boldsymbol{M}^{(+1)} \quad \boldsymbol{M} \quad \boldsymbol{0}_{6\times12}), \qquad \boldsymbol{C}^{(+1,-1)} := (\boldsymbol{M}^{(+1)} \quad \boldsymbol{0}_{6\times4} \quad \boldsymbol{M} \quad \boldsymbol{0}_{6\times8}), \tag{S6}$$

$$\boldsymbol{C}^{(-1,+1)} := (\boldsymbol{M}^{(-1)} \quad \boldsymbol{0}_{6\times8} \quad \boldsymbol{M} \quad \boldsymbol{0}_{6\times4}), \qquad \boldsymbol{C}^{(-1,-1)} := (\boldsymbol{M}^{(-1)} \quad \boldsymbol{0}_{6\times12} \quad \boldsymbol{M})$$

where $\boldsymbol{0}_{m_1\times m_2}$ denotes an $m_1 \times m_2$ matrix whose elements are all zero, and the matrices $\boldsymbol{M}^{(+1)}$, $\boldsymbol{M}^{(-1)}$, and $\boldsymbol{M}$ are:

$$\boldsymbol{M}^{(+1)} := \begin{pmatrix} 1 & 0 & 0 \\ 1 & 1 & 0 \\ 1 & 0 & 0 \\ 1 & 0 & 0 \\ 1 & 0 & 0 \\ 1 & 0 & 0 \end{pmatrix}, \boldsymbol{M}^{(-1)} := \begin{pmatrix} 1 & 0 & 0 \\ 1 & 0 & 0 \\ 1 & 0 & 1 \\ 1 & 0 & 0 \\ 1 & 0 & 0 \\ 1 & 0 & 0 \end{pmatrix}, \boldsymbol{M} := \begin{pmatrix} 0 & 0 & 0 & 0 \\ 0 & 0 & 0 & 0 \\ 1 & 0 & 0 & 0 \\ 0 & 1 & 0 & 0 \\ 0 & 0 & 1 & 0 \\ 0 & 0 & 0 & 1 \end{pmatrix}.$$

## Web Appendix C. Power calculation for a fixed sample size N

Let $\mathbb{P}_M$ denote an empirical mean across $M$ simulated datasets. We estimate power using $\mathbb{P}_M\left[\;\left|\frac{\widehat{\Delta}_Q^{PI}}{\sqrt{\widehat{Var}\left(\widehat{\Delta}_Q^{PI}\right)}}\right| > z_{1-\alpha/2}\;\right]$, calculated as follows:

1. Values for design parameters under the alternative hypothesis ($H_a$) are specified.
2. An appropriate value of the copula dependence parameter $\rho$ (see Web Appendix E for more detail) is selected.
3. A large number, $M,$ of simulated SMART datasets, consisting of $N$ individuals each are generated based on these values; the method of data-generation would follow that described in Section 4.2 in the main manuscript. For each of the $M$ simulated datasets, data from all $N$ individuals will be used to calculate $\widehat{\Delta}_Q^{PI}$ and $\widehat{Var}\left(\widehat{\Delta}_Q^{PI}\right)$.
4. Finally, power is calculated as the proportion of simulated datasets for which the inequality $\left|\frac{\widehat{\Delta}_Q^{PI}}{\sqrt{\widehat{Var}\left(\widehat{\Delta}_Q^{PI}\right)}}\right| > z_{1-\alpha/2}$ holds.

When power calculation is repeated for a grid of sample sizes (e.g., 100, 150, 200, …), to produce a power curve, the sample size needed to attain power, say 0.80, can be determined by selecting the value of $N$ where power first exceeds 0.80.

## Web Appendix D. Approach to generate draws from the multivariate distribution of $\boldsymbol{\theta}_i^g$

Our approach is an application of the method proposed by Madsen and Birkes[6] and consists of three steps. Below, $d_g$ denotes the dimension of $\boldsymbol{\theta}_i^g$, and $\mathbf{0}_{m_1 \times m_2}$ denotes an $m_1 \times m_2$ matrix whose elements are all zero.

1. Generate $n_g$ independent draws from a multivariate standard normal distribution with mean $\mathbf{0}_{d_g \times 1}$ and correlation matrix $\boldsymbol{\Gamma}_{d_g}$, i.e., $\boldsymbol{Z}^{(\boldsymbol{\ell})} \sim MVN\left(\mathbf{0}_{d_g \times 1}, \boldsymbol{\Lambda}_{d_g}\right)$, $\boldsymbol{\ell} = 1,2,\ldots,n_g$.
   - If $\boldsymbol{\Lambda}_{d_g}$ is exchangeable, then $\boldsymbol{\Lambda}_{d_g} = \boldsymbol{I}_{d_g \times d_g} + \rho\left(\mathbf{1}_{d_g \times 1}\mathbf{1}_{d_g \times 1}^T - \boldsymbol{I}_{d_g \times d_g}\right)$ where $\mathbf{1}_{m_1 \times m_2}$ denotes an $m_1 \times m_2$ matrix whose elements are all one and $\boldsymbol{I}_{m_1 \times m_2}$ denotes an $m_1 \times m_2$ identity matrix.
   - If $\boldsymbol{\Lambda}_{d_g}$ is AR1, then the $(\ell, m)$ element of $\boldsymbol{\Lambda}_{d_g}$ is given by $\rho^{|t_\ell - t_m|}$.
2. Denote the vectors $\boldsymbol{Z}^{(\boldsymbol{\ell})}$and $\boldsymbol{U}^{(\boldsymbol{\ell})}$ by $\boldsymbol{Z}^{(\boldsymbol{\ell})} := \begin{pmatrix} Z_{\mathbf{1}}^{(\boldsymbol{\ell})} & \cdots & Z_{d_g}^{(\boldsymbol{\ell})} \end{pmatrix}^{\boldsymbol{T}}$ and $\boldsymbol{U}^{(\boldsymbol{\ell})} := \begin{pmatrix} U_{\mathbf{1}}^{(\boldsymbol{\ell})} & \cdots & U_{d_g}^{(\boldsymbol{\ell})} \end{pmatrix}^{\boldsymbol{T}}$. For each $\ell = 1,2,\ldots,n_g$, generate a new vector $\boldsymbol{U}^{(\boldsymbol{\ell})}$ by applying a transformation using the univariate standard normal CDF to each component of $\boldsymbol{Z}^{(\boldsymbol{\ell})}$. That is, $\begin{pmatrix} U_{\mathbf{1}}^{(\boldsymbol{\ell})} & \cdots & U_{d_g}^{(\boldsymbol{\ell})} \end{pmatrix} = \begin{pmatrix} \phi\left(Z_{\mathbf{1}}^{(\boldsymbol{\ell})}\right) & \cdots & \phi\left(Z_{d_g}^{(\boldsymbol{\ell})}\right) \end{pmatrix}$.
3. For each $\boldsymbol{\ell} = 1,2,\ldots,n_g$, generate a new vector $\boldsymbol{Y}^{(\boldsymbol{\ell})}$ by applying a transformation using the inverse of the CDF of univariate distribution to each component of $\boldsymbol{U}^{(\boldsymbol{\ell})}$. That is, if $\boldsymbol{Y}^{(\boldsymbol{\ell})} := \begin{pmatrix} Y_{\mathbf{1}}^{(\boldsymbol{\ell})} & \cdots & Y_{d_g}^{(\boldsymbol{\ell})} \end{pmatrix}$ and $F_1, \ldots, F_{d_g}$ denotes an appropriate CDF of a univariate distribution (e.g., a NB CDF defined in Section 4.2 of the main manuscript) corresponding to components $1,\ldots,d_g$ respectively, of $\boldsymbol{U}^{(\boldsymbol{\ell})}$, then $\begin{pmatrix} Y_{\mathbf{1}}^{(\boldsymbol{\ell})} & \cdots & Y_{d_g}^{(\boldsymbol{\ell})} \end{pmatrix} = \begin{pmatrix} F_1^{-1}\left(U_{\mathbf{1}}^{(\boldsymbol{\ell})}\right) & \cdots & F_{d_g}^{-1}\left(U_{d_g}^{(\boldsymbol{\ell})}\right) \end{pmatrix}$.

Hence, each $\boldsymbol{Y}^{(\boldsymbol{\ell})}$ is effectively a draw of the potential outcome vector $\boldsymbol{\theta}_i^g$ from its multivariate distribution where the CDFs $F_1, \ldots, F_{d_g}$ are given by Web Table 1 and Web Table 2. The R package *mvtnorm*[6,7] was utilized to draw from a multivariate normal (MVN) distribution.

## Web Appendix E. Approach to estimate the relationship between $\boldsymbol{\rho}$ and $\boldsymbol{\tau_{MAX}}$

Estimating the relationship between $\rho$ and $\tau_{MAX}$ involves generating simulated datasets using the approach described in Section 4 of the main manuscript, and then estimating $Corr\left(Y_{j'}^{s'}, Y_{j''}^{s''}\right)$ for all pairs of ETSs $s'$ and $s''$ and pairs of measurement occasions $j'$ and $j''$ by calculating their average value across all simulated datasets. Specifically, for each value of $\rho$ in a grid (e.g., 0, 0.05, 0.10, 0.15, …), we estimate $\tau_{MAX}$ by using the average of simulated estimates $\hat{\tau}_{MAX}$ calculated using the procedure below.

1. A large number of simulated SMART datasets, $M$, consisting of a large number of individuals each, $N^*$, is generated based values of design parameters specified previously; data-generation would follow that described in Section 4 of the main manuscript, with only the final step on simulating sequential randomizations omitted.
2. For each simulated dataset, we construct the following matrices and estimate $Corr\left(Y_{j'}^{s'}, Y_{j''}^{s''}\right)$ for all pairs of ETSs $s'$ and $s''$ and pairs of measurement occasions $j'$ and $j''$ for pairs of ETSs $s'$ and $s''$ and pairs of measurement occasions $j'$ and $j''$ lying along the path to cell A using the matrix $\boldsymbol{Y}^A$, cell B using the matrix $\boldsymbol{Y}^B$, …, cell F using the matrix $\boldsymbol{Y}^F$; cells A-F are as depicted in the SMART in Figure 1 of the main manuscript.

   - Using individuals belonging to subgroups 1 and 2, construct $\boldsymbol{Y}^A = \begin{pmatrix} \boldsymbol{Y}_1^A \\ \vdots \\ \boldsymbol{Y}_{n_1+n_2}^A \end{pmatrix}$ where $\boldsymbol{Y}_i^A := \begin{pmatrix} Y_{i,1}^{(\cdot)} & \cdots & Y_{i,K}^{(+1)} & Y_{i,K+1}^{(+1,1,0)} & \cdots & Y_{i,T}^{(+1,1,0)} \end{pmatrix}$.

- Using individuals belonging to subgroups 3 and 4, construct $\boldsymbol{Y}^B = \begin{pmatrix} \boldsymbol{Y}_1^B \\ \vdots \\ \boldsymbol{Y}_{n_3+n_4}^B \end{pmatrix}$

  where $\boldsymbol{Y}_i^B := \left(Y_{i,1}^{(\cdot)} \quad \cdots \quad Y_{i,K}^{(+1)} \quad Y_{i,K+1}^{(+1,0,+1)} \quad \cdots \quad Y_{i,T}^{(+1,0,+1)}\right)$.

- Using individuals belonging to subgroups 3 and 4, construct $\boldsymbol{Y}^C = \begin{pmatrix} \boldsymbol{Y}_1^C \\ \vdots \\ \boldsymbol{Y}_{n_3+n_4}^C \end{pmatrix}$

  where $\boldsymbol{Y}_i^C := \left(Y_{i,1}^{(\cdot)} \quad \cdots \quad Y_{i,K}^{(+1)} \quad Y_{i,K+1}^{(+1,0,-1)} \quad \cdots \quad Y_{i,T}^{(+1,0,-1)}\right)$.

- Using individuals belonging to subgroups 1 and 3, construct $\boldsymbol{Y}^D = \begin{pmatrix} \boldsymbol{Y}_1^D \\ \vdots \\ \boldsymbol{Y}_{n_1+n_3}^D \end{pmatrix}$

  where $\boldsymbol{Y}_i^D := \left(Y_{i,1}^{(\cdot)} \quad \cdots \quad Y_{i,K}^{(-1)} \quad Y_{i,K+1}^{(-1,1,0)} \quad \cdots \quad Y_{i,T}^{(-1,1,0)}\right)$.

- Using individuals belonging to subgroups 2 and 4, construct $\boldsymbol{Y}^E = \begin{pmatrix} \boldsymbol{Y}_1^E \\ \vdots \\ \boldsymbol{Y}_{n_2+n_4}^E \end{pmatrix}$

  where $\boldsymbol{Y}_i^E := \left(Y_{i,1}^{(\cdot)} \quad \cdots \quad Y_{i,K}^{(-1)} \quad Y_{i,K+1}^{(-1,0,+1)} \quad \cdots \quad Y_{i,T}^{(-1,0,+1)}\right)$.

- Using individuals belonging to subgroups 2 and 4, construct $\boldsymbol{Y}^F = \begin{pmatrix} \boldsymbol{Y}_1^F \\ \vdots \\ \boldsymbol{Y}_{n_2+n_4}^F \end{pmatrix}$,

  where $\boldsymbol{Y}_i^F := \left(Y_{i,1}^{(\cdot)} \quad \cdots \quad Y_{i,K}^{(-1)} \quad Y_{i,K+1}^{(-1,0,-1)} \quad \cdots \quad Y_{i,T}^{(-1,0,-1)}\right)$.

Now, the quantity $\hat{\tau}_{MAX}$ (defined below) could then be calculated.

$$\hat{\tau}_{MAX} := \max_{s',s'',j',j''} \left\{ \widehat{Corr}\left(Y_{j'}^{s'}, Y_{j''}^{s''}\right) \right\} \tag{S7}$$

When $s' = s''$, the calculation of the maximum in Equation S7 should exclude the terms where $j' = j''$. For example, $\widehat{Corr}\left(Y_{i,2}^{(+1)}, Y_{i,2}^{(+1)}\right) = 1$ is not included in the max.

Let $\hat{\tau}_{MAX}^{\rho}$ denote the average simulated estimate $\hat{\tau}_{MAX}$ corresponding to a particular value of $\rho$ which we calculate using the above-described approach. The value of $\rho$ for which $\hat{\tau}_{MAX}^{\rho}$ is closest to the desired value of $\tau_{MAX}$ will be selected and used to calculate power.

## Web Appendix F. Simulation Study 3

Although the test presented in the main manuscript (Section 2) is expected to perform well asymptotically, it is valuable to use simulations to investigate its performance with finite sample sizes. Simulation Study 3 is designed to evaluate the test's performance by examining the empirical type-I error rate when $\Delta_{EOS} = 0$ and $\Delta_{AUC} = 0$.

As in Simulation Study 1 and 2, we consider a hypothetical two-stage restricted SMART (exemplified by the simplified ENGAGE SMART in Figure 1) where the longitudinal outcome of interest is overdispersed count. There are six monthly measurement occasions (i.e., $T = 6$), randomization of non-responders occurs immediately after the second measurement occasion (i.e., $K = 2$), and response status is defined as $R_i^{(+1)} = I\left(Y_{i,K}^{(+1)} = 0\right)$ and $R_i^{(-1)} = I\left(Y_{i,K}^{(-1)} = 0\right)$. The pair of EDTRs (+1,+1) and (-1,+1) are compared using either difference in end-of-study means, $\Delta_{\text{EOS}} = \mu_6^{(+1,+1)} - \mu_6^{(-1,+1)}$ , or difference in AUC, $\Delta_{\text{AUC}} \approx \sum_{\text{j}=1}^{6-1} \frac{1}{2}\left(\mu_j^{(a_1,a_2^{NR})} + \mu_{j+1}^{(a_1,a_2^{NR})}\right)\left(t_{j+1} - t_j\right)$, each at a desired type-I error rate of $\alpha = 0.05$. We use $p$ and $q$ to denote proportion of responders to first-stage treatment options $a_1 = +1$ and $a_1 = -1$, respectively.

Web Table 6 displays the varying values of parameters used in the scenarios considered in Simulation Study 3. These parameters correspond to increased overdispersion from Scenarios 1-3. Across all scenarios, total sample size $N$, proportion of responders $p$ and $q$, ETS means, and $\rho$ were held constant; an AR1 correlation structure was specified. Three scenarios corresponding to increased values of the NB dispersion parameter were considered, in order to investigate whether higher overdispersion might lead to higher-than-nominal type-I error rates. Finally, we repeated the scenarios described above, but increased the value of $\rho$, in order to investigate whether higher within-person correlation might lead to higher-than-nominal type-I error rates. The magnitudes of $\Delta_{EOS}$ and $\Delta_{\text{AUC}}$ remained unchanged even at increased values of $\rho$. $M = 5000$ Monte Carlo samples were used to calculate empirical type-I error rate.

The results for Simulation Study 3 are summarized in Web Figure 1 (for difference in end-of-study means) and Web Figure 2 (for difference in AUC) where total sample size N (x-axis) is plotted against the empirical type-I error rate (y-axis). Notably, when $\Delta_{AUC} = 0$, empirical type-I error rate ranges between 0.07 to 0.09 at more modest sample sizes (i.e., 200 or less) as overdispersion becomes more extreme (i.e., Scenarios 2 and 3). Otherwise, empirical type-I error rate was either nominal or slightly above nominal (i.e., about 0.05 to 0.07). Higher within-person correlation did not noticeably influence empirical Type-I error rate.

**Web Appendix G. Supplement to Simulation Study 3**

We investigate whether the slightly above nominal or below nominal empirical type-I error rates observed in Simulation Study 3 can be attributed to bias in estimates of $\Delta_{\text{EOS}}$ and $\Delta_{\text{AUC}}$, or bias in estimates of $\sqrt{Var(\hat{\Delta}_{EOS}^{PI})}$ and $\sqrt{Var(\hat{\Delta}_{AUC}^{PI})}$; we utilize values of design parameters identical to Simulation Study 3.

First, bias in estimates of $\Delta_{\text{EOS}}$ and $\Delta_{\text{AUC}}$ were estimated by calculating the average of the difference between the estimated and true value of $\Delta_{\text{EOS}}$ and $\Delta_{\text{AUC}}$. That is, bias is calculated as $\frac{1}{M} \cdot \sum_{k=1}^{M} \left( \hat{\Delta}_{EOS}^{(k)} - \Delta_{\text{EOS}} \right)$ and $\frac{1}{M} \cdot \sum_{k=1}^{M} \left( \hat{\Delta}_{AUC}^{(k)} - \Delta_{\text{AUC}} \right)$, where $\hat{\Delta}_{EOS}^{(k)}$ and $\hat{\Delta}_{AUC}^{(k)}$ denotes the value of $\hat{\Delta}_{EOS}^{PI}$ and $\hat{\Delta}_{AUC}^{PI}$, respectively, at the $k^{th}$ simulated dataset. Note that values of design parameters in Simulation Study 3 were specified such that $\Delta_{EOS} = 0$ and $\Delta_{\text{AUC}} = 0$. As Web Figure 3 shows, estimates $\hat{\Delta}_{EOS}^{PI}$ are unbiased. On the other hand, as Web Figure 4 shows, $\hat{\Delta}_{AUC}^{PI}$ exhibits slight bias which attenuates as total sample size N is increased.

Second, bias in estimates of $\sqrt{Var(\hat{\Delta}_{EOS}^{PI})}$ and $\sqrt{Var(\hat{\Delta}_{AUC}^{PI})}$ was estimated by calculating the average of the difference between the estimated value of $\sqrt{Var(\hat{\Delta}_{EOS}^{PI})}$ and $\sqrt{Var(\hat{\Delta}_{AUC}^{PI})}$, and the empirical standard error of $\hat{\Delta}_{EOS}^{PI}$ and $\hat{\Delta}_{AUC}^{PI}$, respectively. Here, empirical standard error is calculated as the square root of the variance of $\hat{\Delta}_{EOS}^{PI}$ and $\hat{\Delta}_{AUC}^{PI}$ across all simulated datasets. That is, bias is calculated as $\frac{1}{M} \cdot \sum_{k=1}^{M} \left(\hat{\sigma}_{EOS}^{(k)} - \sigma_{\mathrm{EOS}}\right)$ and $\frac{1}{M} \cdot \sum_{k=1}^{M} \left(\hat{\sigma}_{AUC}^{(k)} - \sigma_{\mathrm{AUC}}\right)$, where $\hat{\sigma}_{EOS}^{(k)}$ and $\hat{\sigma}_{AUC}^{(k)}$ denotes the value of $\sqrt{\widehat{Var}(\hat{\Delta}_{EOS}^{PI})}$ and $\sqrt{\widehat{Var}(\hat{\Delta}_{AUC}^{PI})}$, respectively, at the $k^{th}$ simulated dataset; $\sigma_{\mathrm{EOS}}$ and $\sigma_{\mathrm{AUC}}$ denotes the empirical standard error of $\hat{\Delta}_{EOS}^{PI}$ and $\hat{\Delta}_{AUC}^{PI}$, respectively. As Web Figure 5 shows, estimates of standard errors for differences in end-of-study means are unbiased. On the other hand, as Web Figure 6 shows, estimates of $\sqrt{Var(\hat{\Delta}_{AUC}^{PI})}$ exhibit a pronounced *downward* bias when total sample size N is 200 or less; the bias attenuated as N was increased.

## Web Appendix H. Simulation Study 4

We investigate whether power is sensitive to violation of the working assumption made with respect to $\eta$, namely that $\eta = \frac{\rho}{2}$.

As in Simulation Study 1 and 2, we consider a hypothetical two-stage restricted SMART (exemplified by the simplified ENGAGE SMART in Figure 1) where the longitudinal outcome of interest is overdispersed count. There are six monthly measurement occasions (i.e., $T = 6$), randomization of non-responders occurs immediately after the second measurement occasion (i.e., $K = 2$), and response status is defined as $R_i^{(+1)} = I\left(Y_{i,K}^{(+1)} = 0\right)$ and $R_i^{(-1)} = I\left(Y_{i,K}^{(-1)} = 0\right)$. The pair of EDTRs (+1,+1) and (-1,+1) are compared using either difference in end-of-study means, $\Delta_{\text{EOS}} = \mu_6^{(+1,+1)} - \mu_6^{(-1,+1)}$ , or difference in AUC, $\Delta_{\text{AUC}} \approx \sum_{j=1}^{6-1} \frac{1}{2}\left(\mu_j^{(a_1,a_2^{NR})} + \mu_{j+1}^{(a_1,a_2^{NR})}\right)\left(t_{j+1} - t_j\right)$, each at a desired type-I error rate of $\alpha = 0.05$. We use $p$ and $q$ to denote proportion of responders to first-stage treatment options $a_1 = +1$ and $a_1 = -1$, respectively.

Ten scenarios identical to those described in Simulation Study 1 were considered, except that the total sample size N was fixed to 500 and $\rho$ was fixed to 0.6. Within each scenario, the working assumption was violated by calculating power when $\eta$ was set to 0, 0.05, 0.10, 0.15, 0.20, …, 0.45. $M = 5000$ Monte Carlo samples were used to calculate power. Throughout, an AR1 correlation structure was specified.

We note that investigating sensitivity of violations to this working assumption is only possible at pairs of values of $\rho$ and $\eta$ for which each of the $\mathbf{\Lambda}_d$'s is positive definite. When $\rho$ was fixed to 0.6, values of $\eta$ equal to 0.50, 0.55, 0.60, 0.65, …, 0.95, 1.00 will result in at least one of the $\mathbf{\Lambda}_d$'s being non-positive-definite.

The results for Simulation Study 4 are summarized in Web Figure 7 where $\eta$ (x-axis) is plotted against power (y-axis). More specifically, results for scenario 1 (i.e., when $\Delta_{\text{EOS}}$=0.28 [left panel]; when $\Delta_{\text{AUC}}$= 0.7 [right panel]) are displayed as solid dots on the bottom-most dashed horizontal line, results for scenario 2 (i.e., when $\Delta_{\text{EOS}}$=0.56 [left panel]; when $\Delta_{\text{AUC}}$= 1.41 [right panel]) are displayed as solid dots on the 2nd dashed horizontal line from the bottom. Analogously, results for scenarios 3-10 are displayed as solid dots on the 3rd through 10th dashed horizontal line from the bottom. The results show that across the ten scenarios, power is not sensitive to the actual value of $\eta$.

## Web Appendix I. Simulation Study 5

We investigate whether power is sensitive to violation of the working assumption made with respect to subgroup 4, namely that the number of individuals in subgroup 4 (that is, $n_4$) is equal to the minimum number of non-responders to either of the 2 initial treatments.

As in Simulation Study 1 and 2, we consider a hypothetical two-stage restricted SMART (exemplified by the simplified ENGAGE SMART in Figure 1) where the longitudinal outcome of interest is overdispersed count. There are six monthly measurement occasions (i.e., $T = 6$), randomization of non-responders occurs immediately after the second measurement occasion (i.e., $K = 2$), and response status is defined as $R_i^{(+1)} = I\left(Y_{i,K}^{(+1)} = 0\right)$ and $R_i^{(-1)} = I\left(Y_{i,K}^{(-1)} = 0\right)$. The pair of EDTRs (+1,+1) and (-1,+1) are compared using either difference in end-of-study means, $\Delta_{\text{EOS}} = \mu_6^{(+1,+1)} - \mu_6^{(-1,+1)}$ , or difference in AUC, $\Delta_{\text{AUC}} \approx \sum_{\text{j}=1}^{6-1} \frac{1}{2}\left(\mu_j^{(a_1,a_2^{NR})} + \mu_{j+1}^{(a_1,a_2^{NR})}\right)\left(t_{j+1} - t_j\right)$, each at a desired type-I error rate of $\alpha = 0.05$. We use $p$ and $q$ to denote proportion of responders to first-stage treatment options $a_1 = +1$ and $a_1 = -1$, respectively.

Ten scenarios identical to those described in Simulation Study 1 were considered, except that the total sample size N was fixed to 500 and $\rho$ was fixed to 0.6. Within each scenario, the working assumption was violated by calculating power when $n_4$ was set to 100, 150, 200, 250, 300 (i.e., the maximum possible value of $n_4$). $M = 5000$ Monte Carlo samples were used to calculate power. Throughout, an AR1 correlation structure was specified. The results for Simulation Study 5 are summarized in Web Figure 8 where $n_4$ (x-axis) is plotted against power (y-axis). More specifically, results for scenario 1 (i.e., when $\Delta_{\text{EOS}}$=0.28 [left panel]; when $\Delta_{\text{AUC}}$= 0.7 [right panel]) are displayed as solid dots on the bottom-most dashed horizontal line, results for scenario 2 (i.e., when $\Delta_{\text{EOS}}$=0.56 [left panel]; when $\Delta_{\text{AUC}}$= 1.41 [right panel]) are displayed as solid dots on the 2nd dashed horizontal line from the bottom. Analogously, results for scenarios 3-10 are displayed as solid dots on the 3rd through 10th dashed horizontal line from the bottom. The results show that across the ten scenarios, power is not sensitive to the actual value of $n_4$.

## Web Appendix J. Estimating sample size when missing data is anticipated in a planned SMART

An additional step in the data generating process (described in Section 4 of the main manuscript) can be included to induce missing completely at random or missing at random (MCAR or MAR)[7, 8] values in each of the $M$ simulated datasets. Specifically, for each of the $M$ simulated datasets (say, $M = 5000$), MCAR values in the longitudinal count outcome can be subsequently generated by regarding each observed outcome $Y_{i,j}$ as missing when $\mathcal{O}_{i,j} = 0$ and not missing when $\mathcal{O}_{i,j} = 1$, where $\mathcal{O}_{i,j} \sim$ Bernoulli$(\mathcal{p})$ and $\mathcal{p} \in (0,1)$. In contrast, for each of the $M$ simulated datasets, MAR values in the longitudinal outcome can subsequently be generated by first specifying a logistic regression model for the (conditional) probability that $Y_{i,j}$ is not missing at measurement occasion $j$ (which we denote by $\mathcal{p}_j$) in terms of variables observed prior to measurement occasion $j$, and then, regarding each observed outcome $Y_{i,j}$ as missing when $\mathcal{O}_{i,j} = 0$ and not missing when $\mathcal{O}_{i,j} = 1$, where $\mathcal{O}_{i,j} \sim$ Bernoulli$\left(\mathcal{p}_j\right)$ and $\mathcal{p}_j \in (0,1)$. As a concrete example, consider a hypothetical two-stage restricted SMART (exemplified by the simplified ENGAGE SMART in Figure 1). In this case, one possible model for a MAR missing data mechanism might be $logit\left(\mathcal{p}_j\right) = \gamma_0 + \gamma_1 X + \gamma_2 A_1 I(1 < j \le T) + \gamma_3 X A_1 I(1 < j \le T)$ for all $j = 1, \ldots, T$, where $A_1 \in \{-1, +1\}$ is an indicator for first-stage randomization assignment and $X$ represents a binary baseline covariate (e.g., which may represent gender, race, etc.)

defined as a deterministic function of the outcome at the first measurement occasion, e.g., $X = I(Y_{i,j} < \omega)$ for some cutoff $\omega > 0$. Modeling more complex dependence between $X$ and missingness may require eliciting additional parameters from domain experts; this is an area for future work.

To deal with missing data in each of the $M$ simulated datasets, multiple imputation can be employed.[9-11] Standard corrections (e.g., see pp. 86-87 in Little and Rubin[9]) can then be used to adjust our proposed test (see Section 2 of main manuscript) to account for uncertainty from multiple imputation. Finally, power can then be estimated by the proportion of times out of $M$ for which the null hypothesis is rejected using this adjusted test. That is, if $S$ denotes the total number of imputed datasets associated with each of the $M$ simulated datasets, then power can then be estimated by $\mathbb{P}_M\left\{I\left(\left|\frac{\overline{\Delta}_Q^S}{\overline{\sigma}_Q^S}\right| > z_{1-\alpha/2}\right)\right\}$ where the terms within the indicator function are given by the equations below.

$$\overline{\Delta}_Q^S := \frac{1}{S}\sum_{k=1}^{S} \widehat{\Delta}_Q^{(k)}$$

$$\overline{\sigma}_Q^S := \sqrt{W + \left(1 + \frac{1}{S}\right) \cdot B}$$

$$W := \frac{1}{S}\sum_{k=1}^{S}\left(\hat{\sigma}_Q^{(k)}\right)^2 \text{ and } B := \frac{1}{S-1} \cdot \sum_{k=1}^{S}\left(\widehat{\Delta}_Q^{(k)} - \overline{\Delta}_Q^S\right)^2$$

In these equations, $\widehat{\Delta}_Q^{(k)}$ and $\hat{\sigma}_Q^{(k)}$ denote the value of $\hat{\Delta}_Q^{PI}$ and $\sqrt{\widehat{Var}(\hat{\Delta}_{EOS}^{PI})}$, respectively, estimated using the $k^{th}$ imputed dataset ($k = 1, \ldots, S$).

## Web Appendix K. Software and Computation Time

The R package[12] *geeM*[13] was utilized in this manuscript's implementation of the IPWR estimator; the R package *mvtnorm*[14] was utilized to obtain samples from a MVN distribution; the R package *rootSolve* was used for root-finding.[15, 16]

Computation time to complete calculation of power for one given set of parameters in Simulation Studies 1-5 did not go beyond 900 seconds (15 minutes), and it decreased with smaller sample sizes. Actual computation time to complete calculation of power for particular sets of parameters considered displayed in a column named 'elapsed.secs' within files named 'power.csv', along with R code[12] implementing the simulation studies, is in the file CountSMART-2.0.0.zip.

## Web Figures

(Web Figures begin on the following page)

**Web Figure 1:** Results of Simulation Study 3 in Web Appendix F – Empirical Type-I Error Rate when $\Delta_{EOS} = 0$.

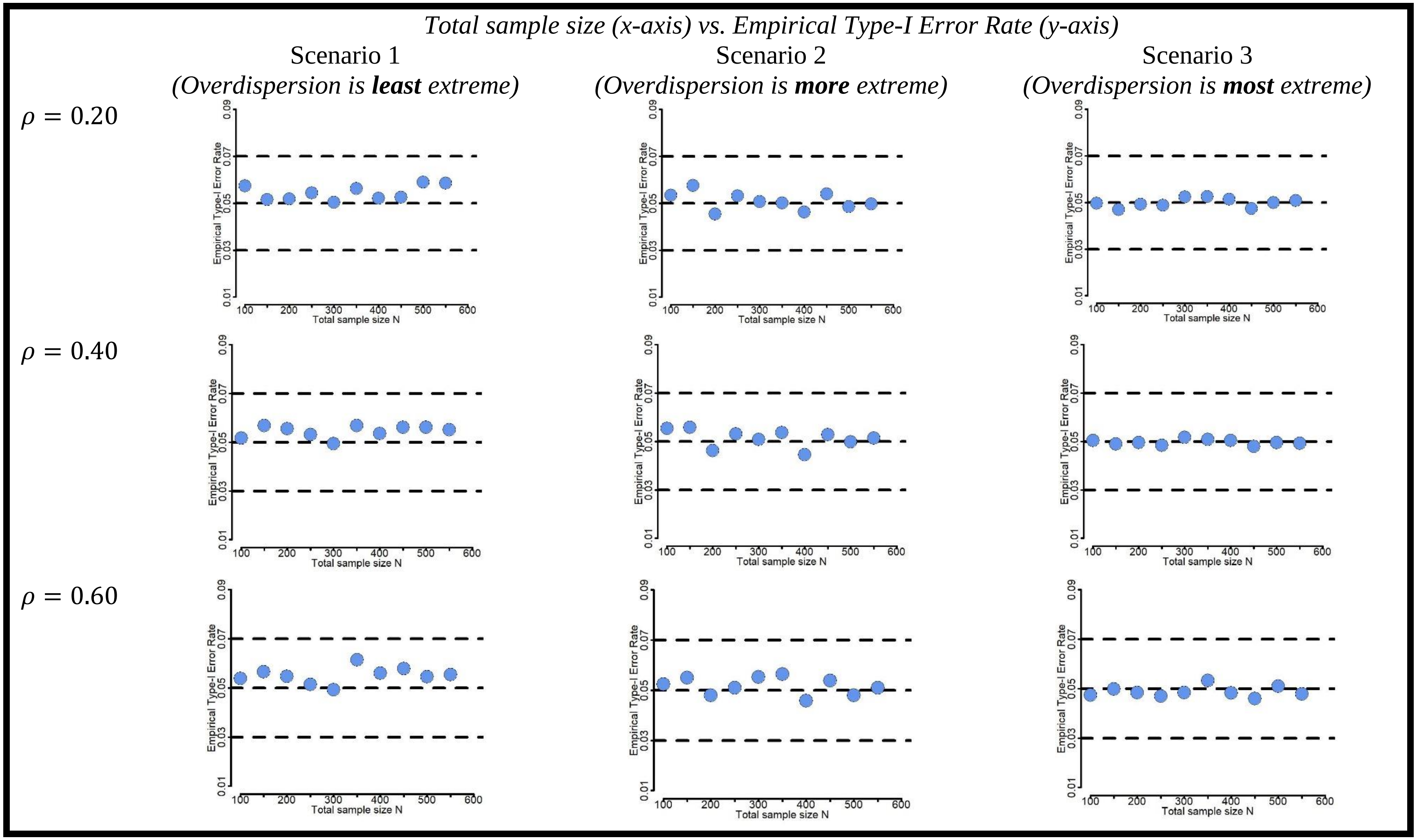

**Web Figure 2:** Results of Simulation Study 3 in Web Appendix F – Empirical Type-I Error Rate when $\Delta_{AUC} = 0$.

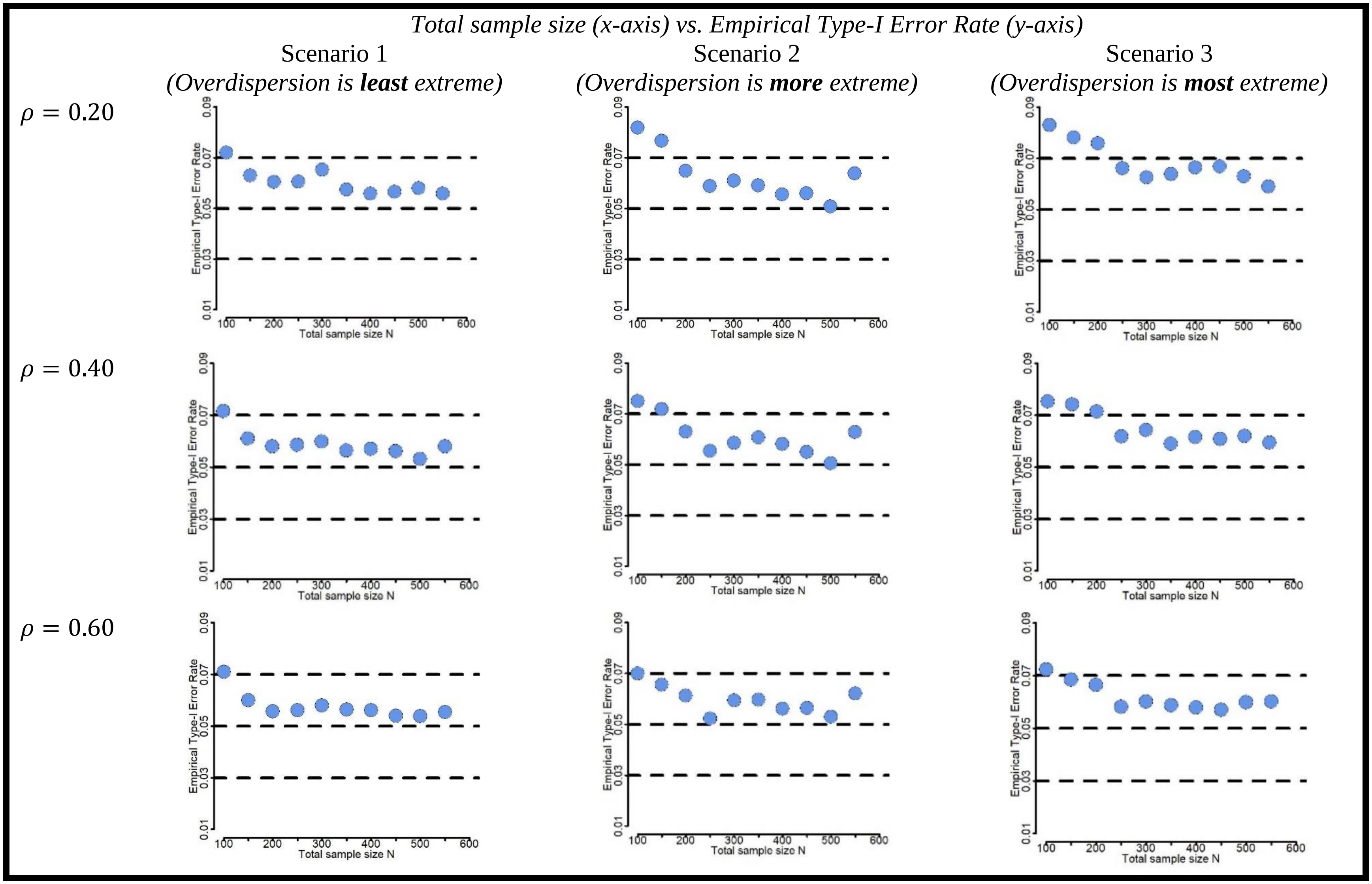

**Web Figure 3:** Results of Supplement to Simulation Study 3 in Web Appendix G – Bias when $\Delta_{EOS} = 0$.

*Total sample size (x-axis) vs. Empirical Bias in estimates* of $\Delta^{PI}_{EOS}$ *(y-axis)*

| | Scenario 1 *(Overdispersion is **least** extreme)* | Scenario 2 *(Overdispersion is **more** extreme)* | Scenario 3 *(Overdispersion is **most** extreme)* |
|---|---|---|---|
| $\rho = 0.20$ | | | |
| $\rho = 0.40$ | | | |
| $\rho = 0.60$ | | | |

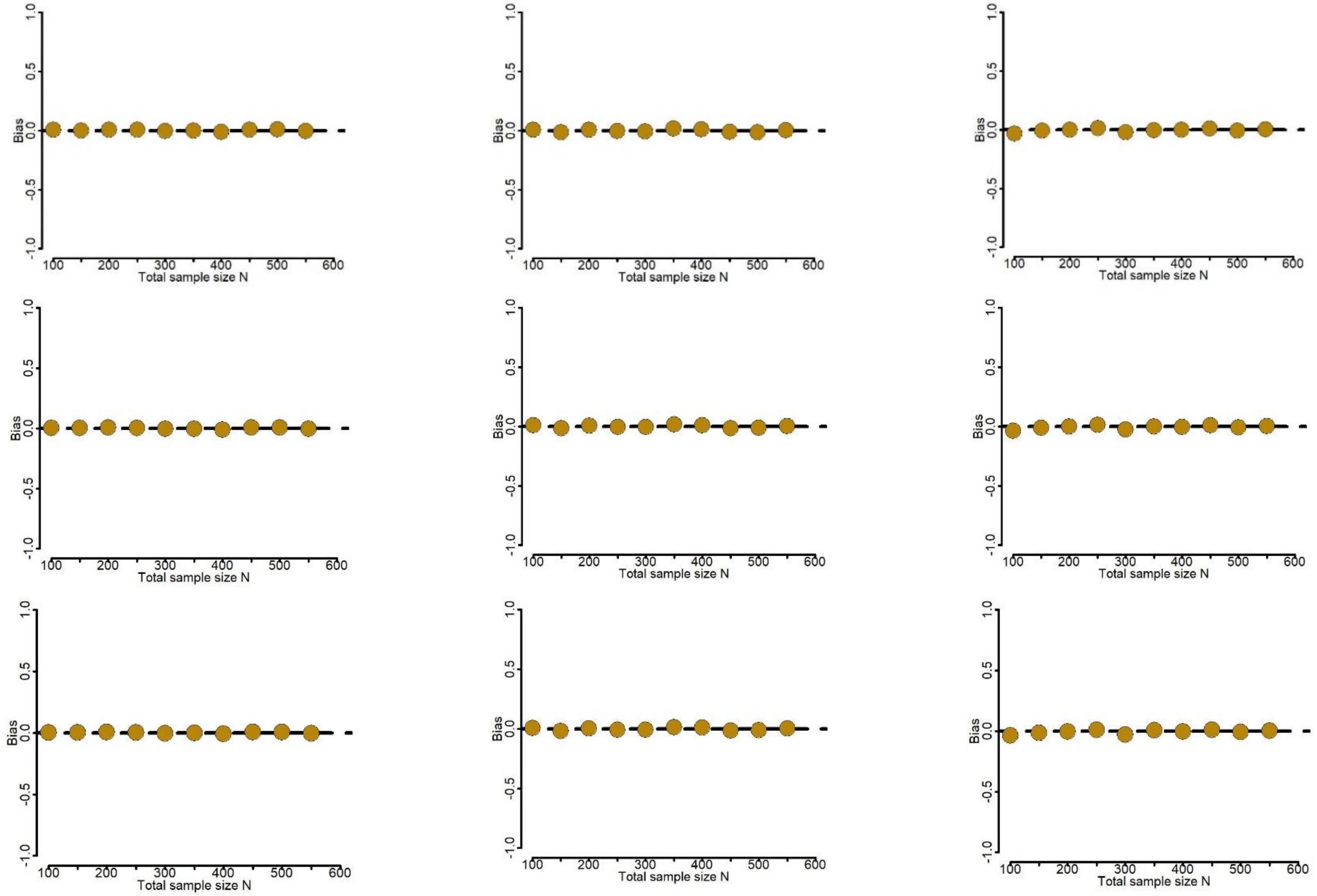

**Web Figure 4:** Results of Supplement to Simulation Study 3 in Web Appendix G – Bias when $\Delta_{AUC} = 0$.

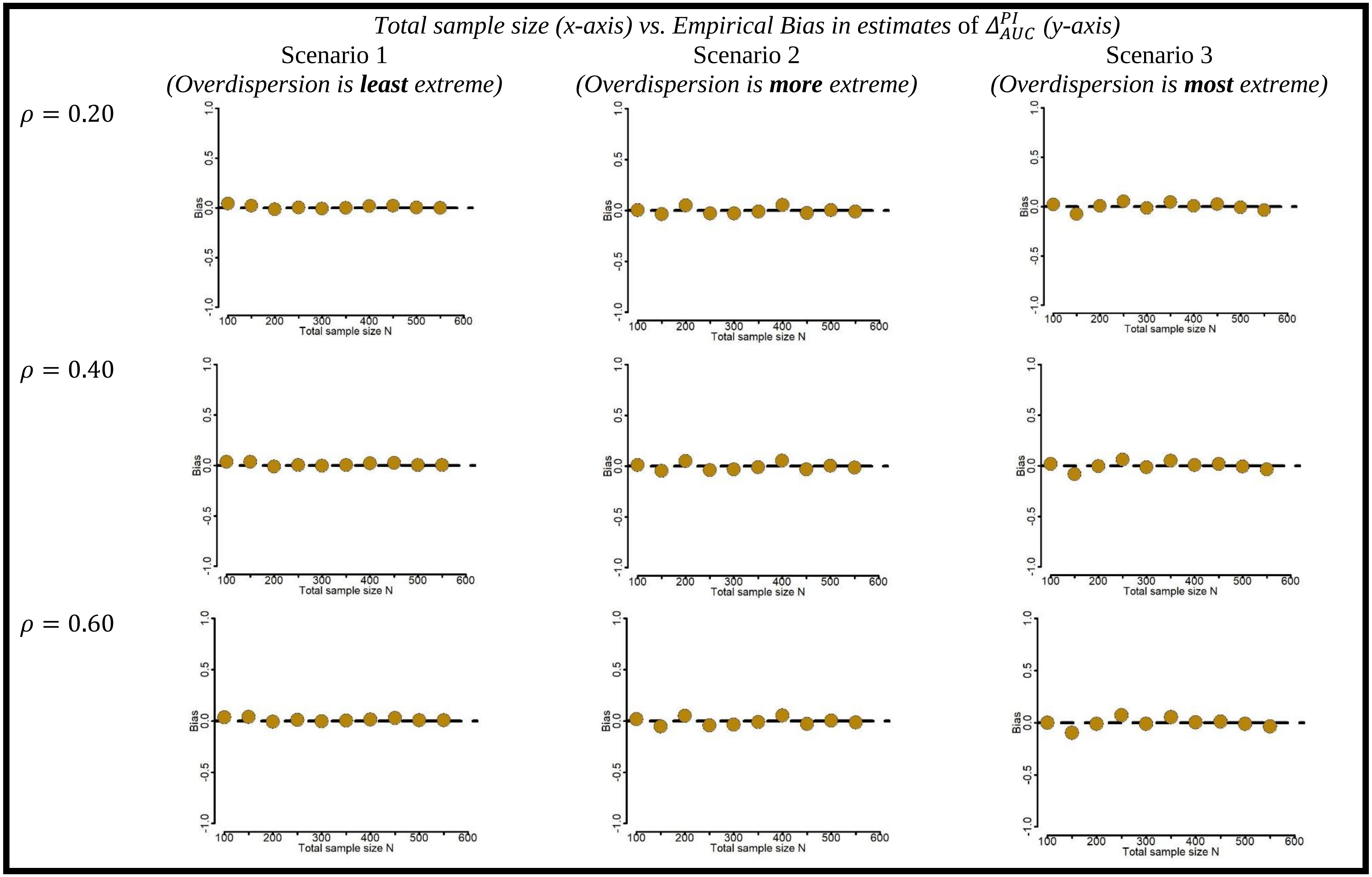

**Web Figure 5:** Results of Supplement to Simulation Study 3 in Web Appendix G – Bias when $\Delta_{EOS} = 0$.

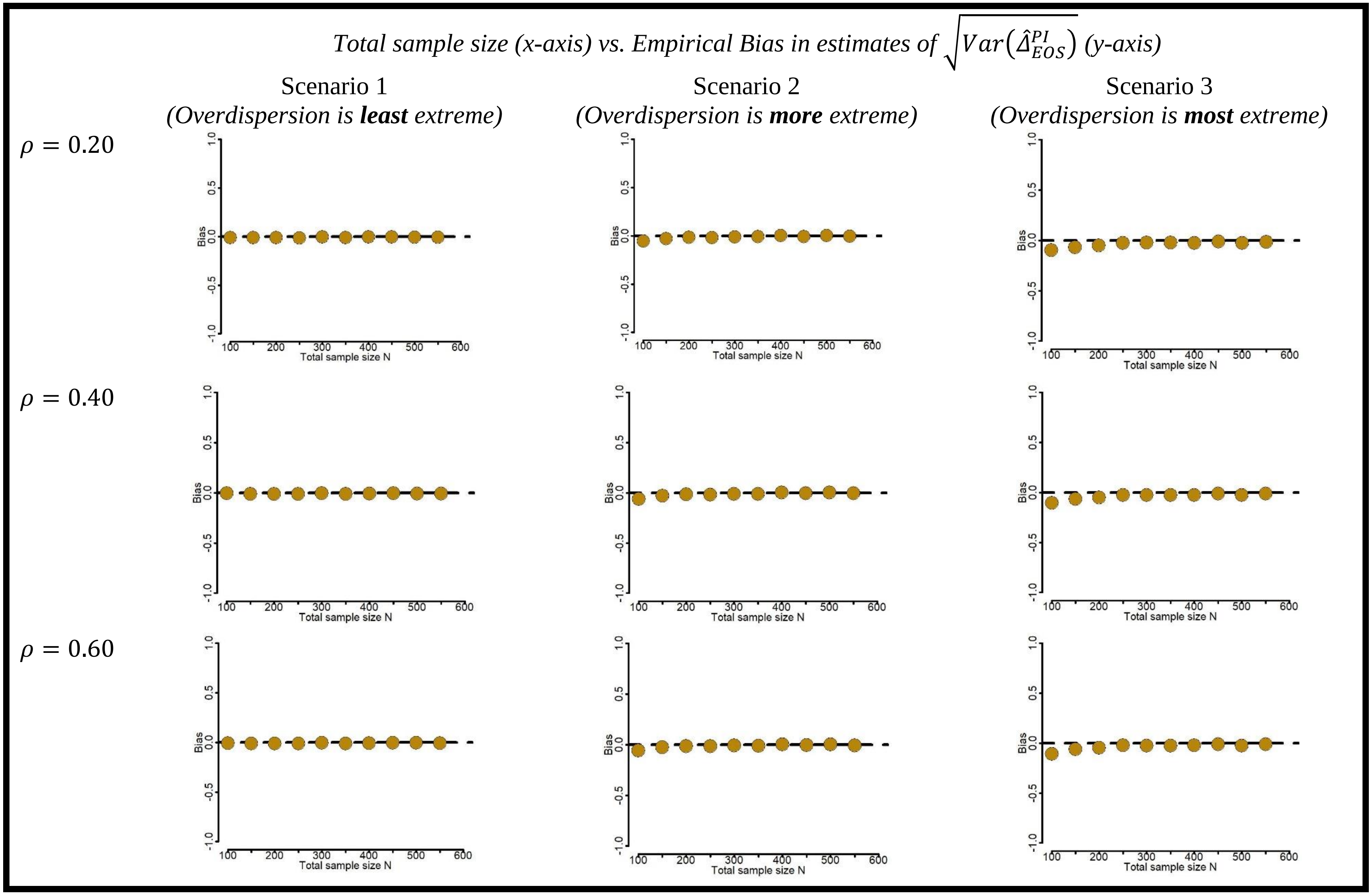

**Web Figure 6:** Results of Supplement to Simulation Study 3 in Web Appendix G – Bias when $\Delta_{AUC} = 0$.

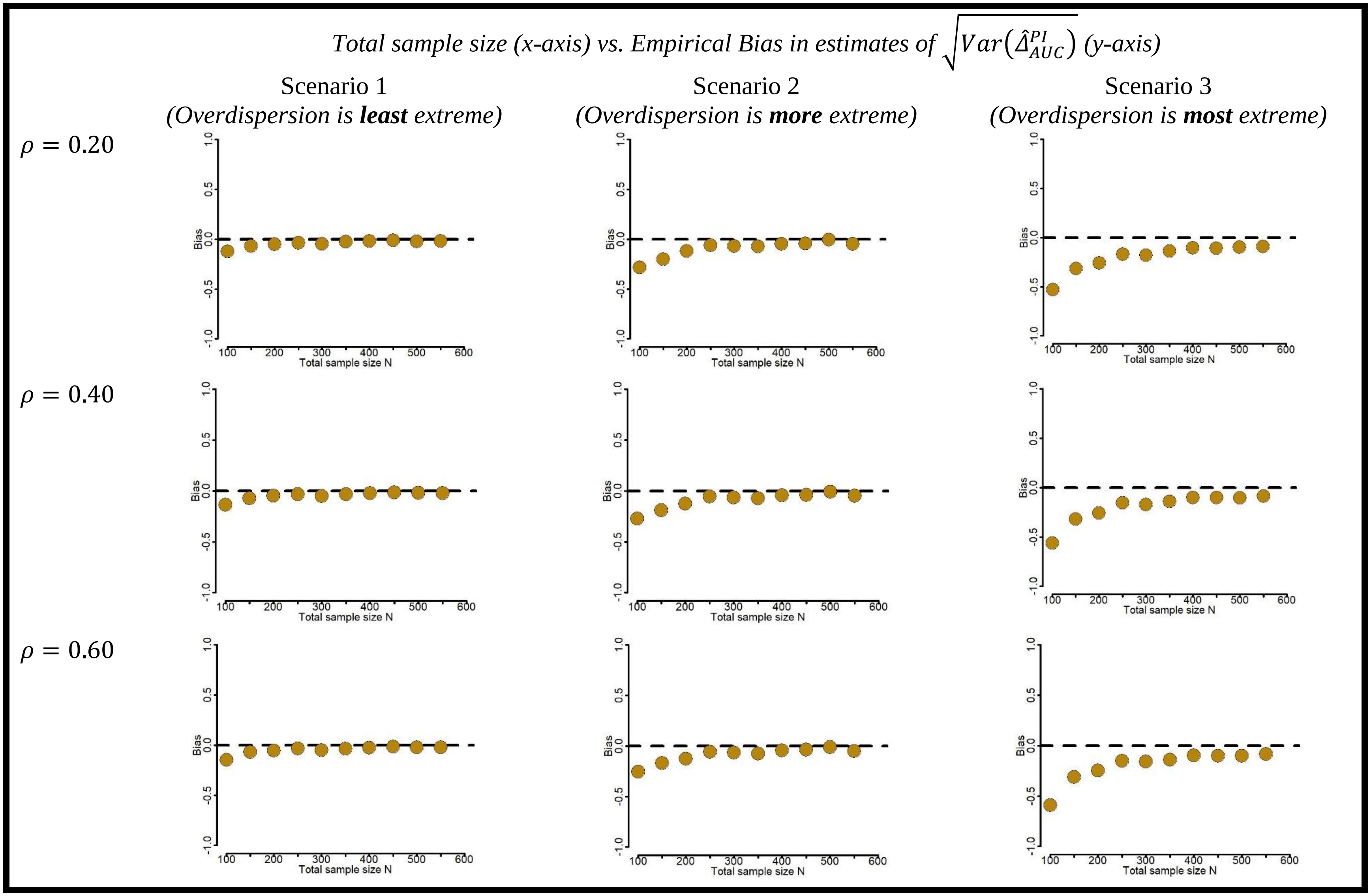

**Web Figure 7:** Results of Simulation Study 4 in Web Appendix H – Sensitivity of power to violations in working assumption on $\eta$

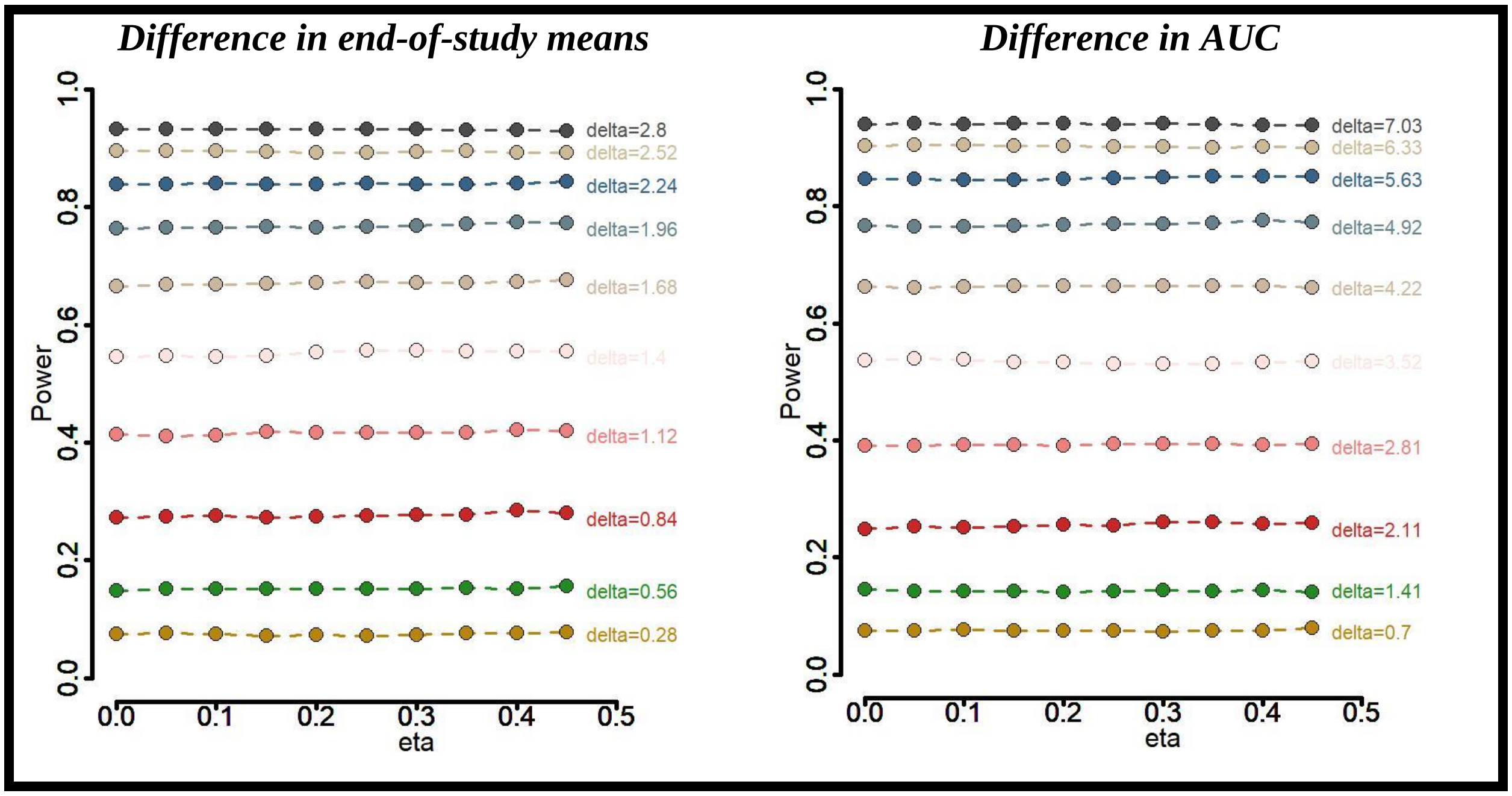

**Web Figure 8:** Results of Simulation Study 5 in Web Appendix I – Sensitivity of power to violations in working assumption on $n_4$

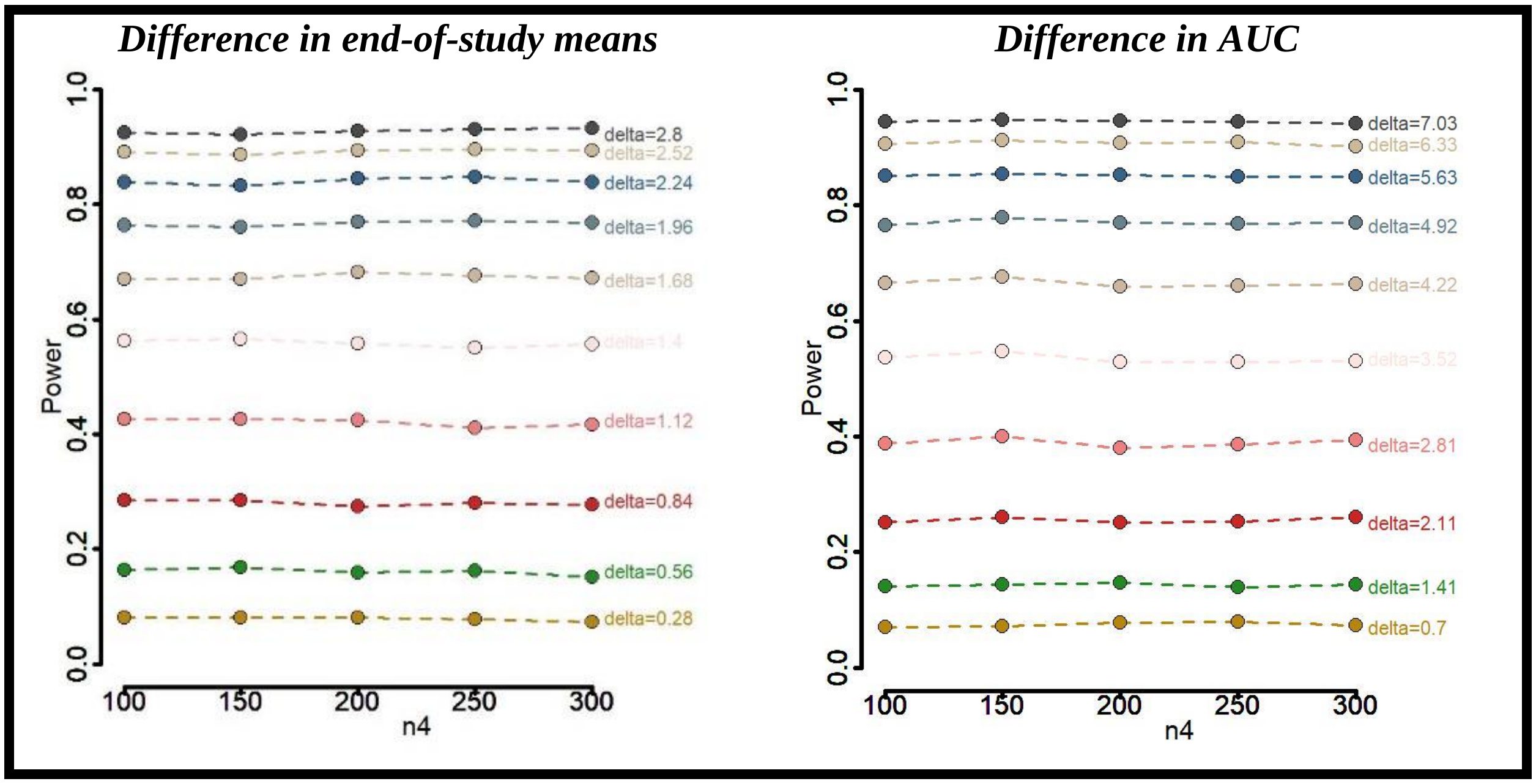

**Web Figure 9:** Results of Simulation Study 1 when an AR1 structure is utilized in the specification of the latent MVN-distributed variable.

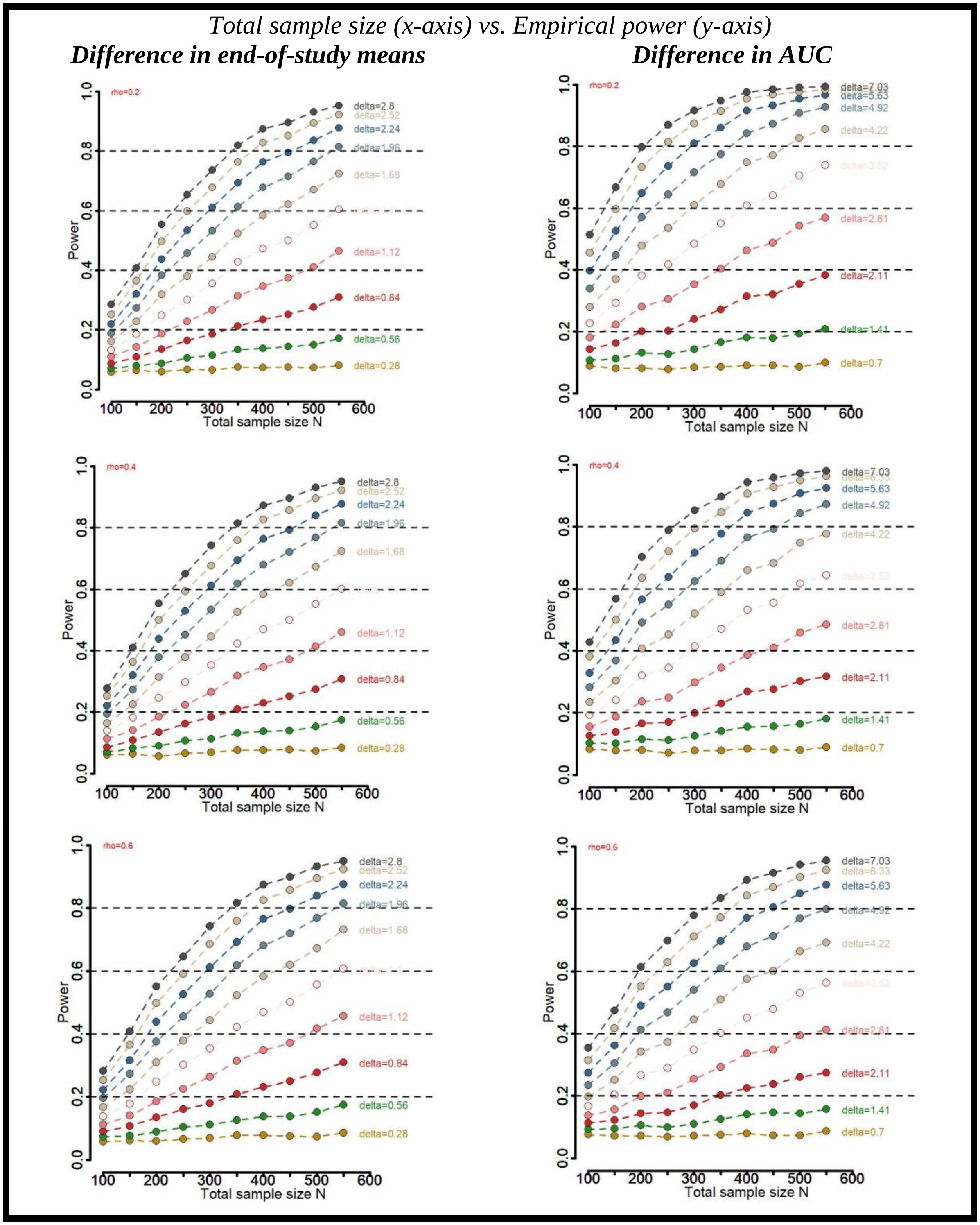

**Web Figure 10:** Results of Simulation Study 1 when an exchangeable structure is utilized in the specification of the latent MVN-distributed variable.

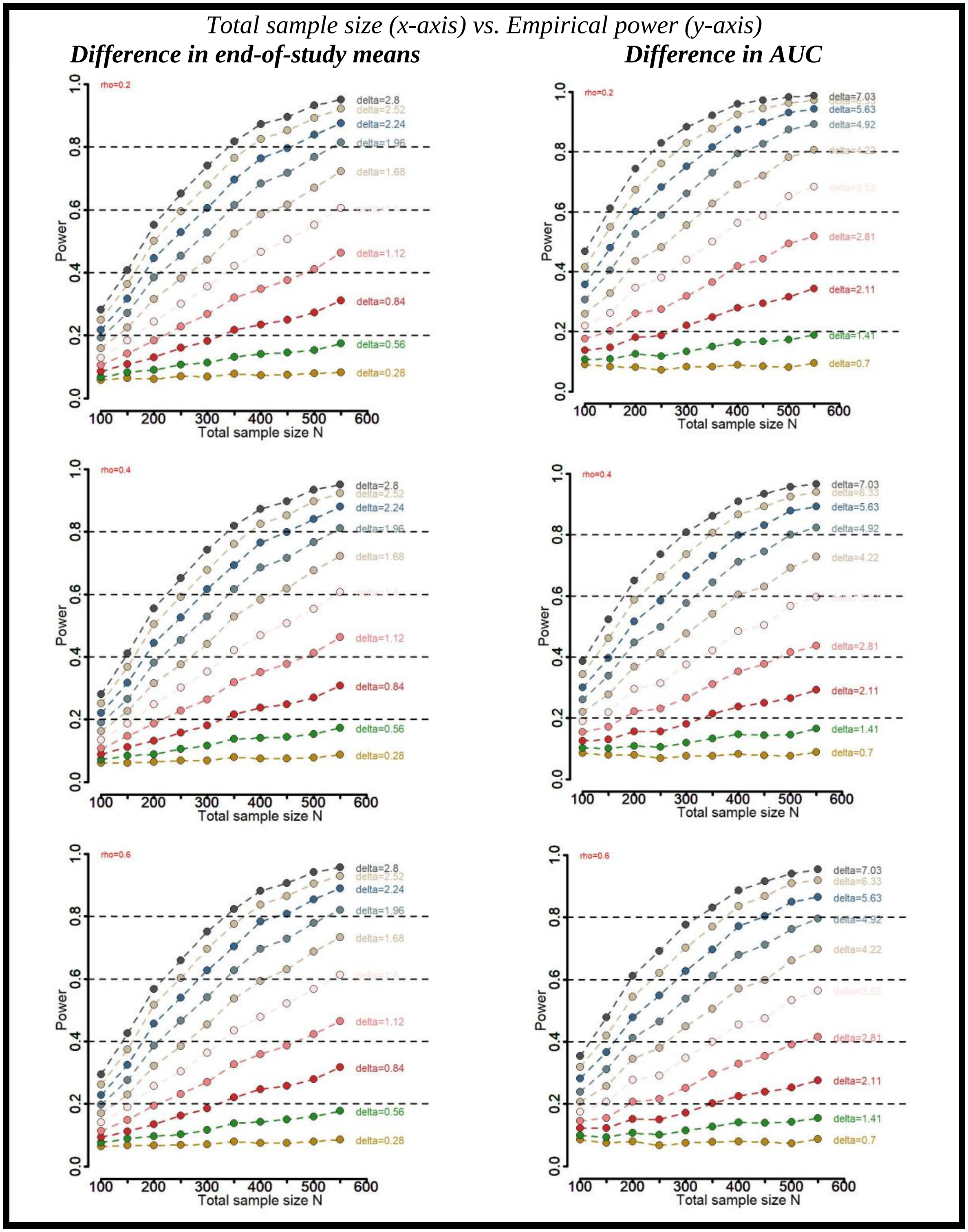

**Web Figure 11:** Results of Simulation Study 2 when an AR1 structure is utilized in the specification of the latent MVN-distributed variable.

*A display of empirical within-person correlation among simulated count outcomes when $\rho = 0.6$ in Scenario 10, averaged across $M = 5000$ Monte Carlo samples. Panels (a), (b), (c), (d), (e), (f) display correlation of count outcomes was calculated among those which correspond to ETS lying on the path leading to cell A, B, C, D, E, F in Figure 1, respectively.*

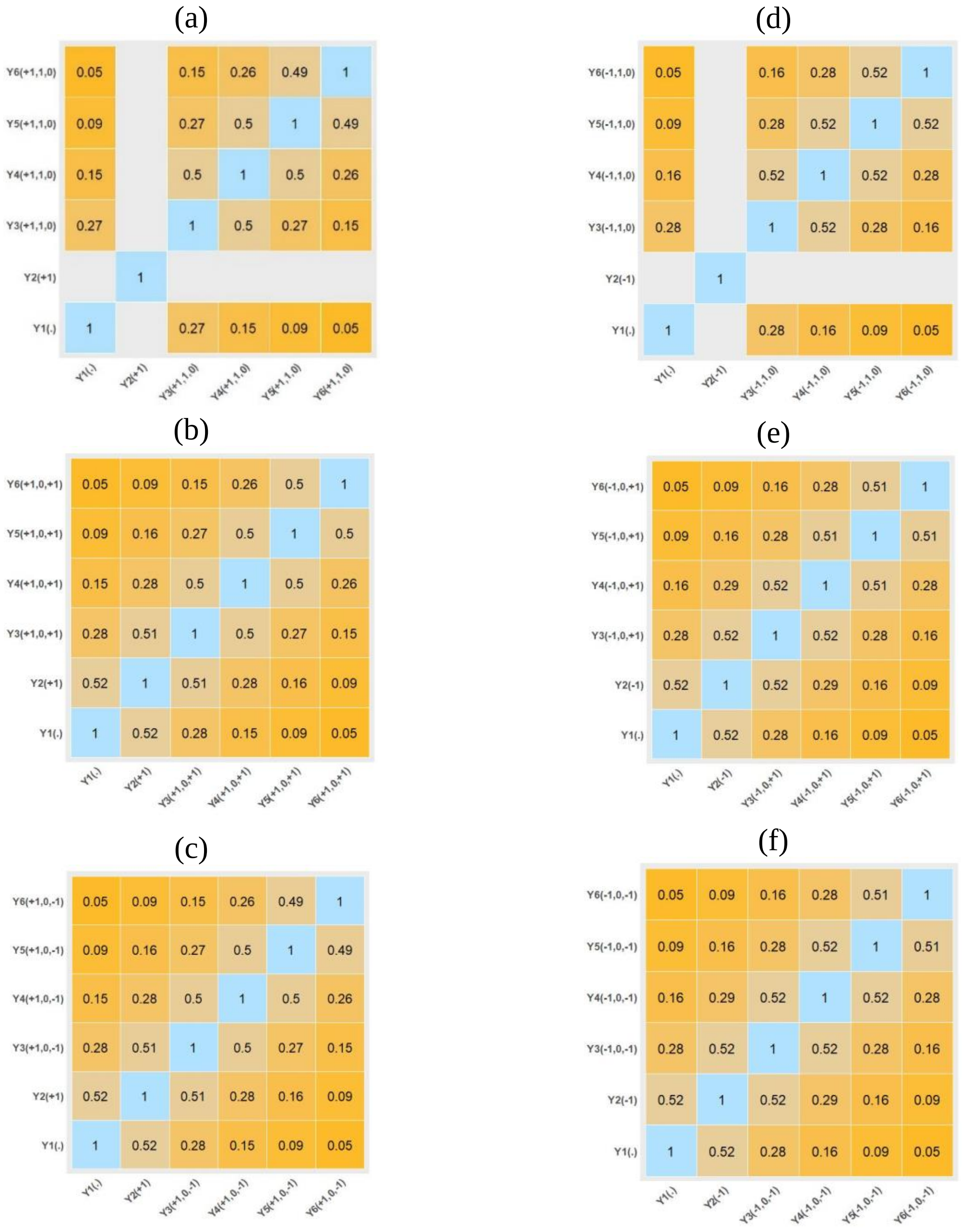

**Web Figure 12:** Results of Simulation Study 2 when an exchangeable structure is utilized in the specification of the latent MVN-distributed variable.

*A display of empirical within-person correlation among simulated count outcomes when $\rho = 0.6$ in Scenario 10, averaged across $M = 5000$ Monte Carlo samples. Panels (a), (b), (c), (d), (e), (f) display correlation of count outcomes was calculated among those which correspond to ETS lying on the path leading to cell A, B, C, D, E, F in Figure 1, respectively.*

(a)

| | Y1(.) | Y2(+1) | Y3(+1,1,0) | Y4(+1,1,0) | Y5(+1,1,0) | Y6(+1,1,0) |
|---|---|---|---|---|---|---|
| Y6(+1,1,0) | 0.5 | | 0.5 | 0.49 | 0.49 | 1 |
| Y5(+1,1,0) | 0.51 | | 0.5 | 0.5 | 1 | 0.49 |
| Y4(+1,1,0) | 0.51 | | 0.5 | 1 | 0.5 | 0.49 |
| Y3(+1,1,0) | 0.51 | | 1 | 0.5 | 0.5 | 0.5 |
| Y2(+1) | | 1 | | | | |
| Y1(.) | 1 | | 0.51 | 0.51 | 0.51 | 0.5 |

(d)

| | Y1(.) | Y2(-1) | Y3(-1,1,0) | Y4(-1,1,0) | Y5(-1,1,0) | Y6(-1,1,0) |
|---|---|---|---|---|---|---|
| Y6(-1,1,0) | 0.52 | | 0.52 | 0.52 | 0.52 | 1 |
| Y5(-1,1,0) | 0.52 | | 0.52 | 0.52 | 1 | 0.52 |
| Y4(-1,1,0) | 0.52 | | 0.52 | 1 | 0.52 | 0.52 |
| Y3(-1,1,0) | 0.52 | | 1 | 0.52 | 0.52 | 0.52 |
| Y2(-1) | | 1 | | | | |
| Y1(.) | 1 | | 0.52 | 0.52 | 0.52 | 0.52 |

(b)

| | Y1(.) | Y2(+1) | Y3(+1,0,+1) | Y4(+1,0,+1) | Y5(+1,0,+1) | Y6(+1,0,+1) |
|---|---|---|---|---|---|---|
| Y6(+1,0,+1) | 0.5 | 0.5 | 0.5 | 0.49 | 0.5 | 1 |
| Y5(+1,0,+1) | 0.51 | 0.51 | 0.5 | 0.5 | 1 | 0.5 |
| Y4(+1,0,+1) | 0.51 | 0.51 | 0.5 | 1 | 0.5 | 0.49 |
| Y3(+1,0,+1) | 0.51 | 0.51 | 1 | 0.5 | 0.5 | 0.5 |
| Y2(+1) | 0.52 | 1 | 0.51 | 0.51 | 0.51 | 0.5 |
| Y1(.) | 1 | 0.52 | 0.51 | 0.51 | 0.51 | 0.5 |

(e)

| | Y1(.) | Y2(-1) | Y3(-1,0,+1) | Y4(-1,0,+1) | Y5(-1,0,+1) | Y6(-1,0,+1) |
|---|---|---|---|---|---|---|
| Y6(-1,0,+1) | 0.52 | 0.52 | 0.51 | 0.51 | 0.51 | 1 |
| Y5(-1,0,+1) | 0.52 | 0.52 | 0.51 | 0.51 | 1 | 0.51 |
| Y4(-1,0,+1) | 0.52 | 0.52 | 0.52 | 1 | 0.51 | 0.51 |
| Y3(-1,0,+1) | 0.52 | 0.52 | 1 | 0.52 | 0.51 | 0.51 |
| Y2(-1) | 0.52 | 1 | 0.52 | 0.52 | 0.52 | 0.52 |
| Y1(.) | 1 | 0.52 | 0.52 | 0.52 | 0.52 | 0.52 |

(c)

| | Y1(.) | Y2(+1) | Y3(+1,0,-1) | Y4(+1,0,-1) | Y5(+1,0,-1) | Y6(+1,0,-1) |
|---|---|---|---|---|---|---|
| Y6(+1,0,-1) | 0.5 | 0.51 | 0.5 | 0.49 | 0.49 | 1 |
| Y5(+1,0,-1) | 0.51 | 0.51 | 0.5 | 0.5 | 1 | 0.49 |
| Y4(+1,0,-1) | 0.51 | 0.51 | 0.5 | 1 | 0.5 | 0.49 |
| Y3(+1,0,-1) | 0.51 | 0.51 | 1 | 0.5 | 0.5 | 0.5 |
| Y2(+1) | 0.52 | 1 | 0.51 | 0.51 | 0.51 | 0.51 |
| Y1(.) | 1 | 0.52 | 0.51 | 0.51 | 0.51 | 0.5 |

(f)

| | Y1(.) | Y2(-1) | Y3(-1,0,-1) | Y4(-1,0,-1) | Y5(-1,0,-1) | Y6(-1,0,-1) |
|---|---|---|---|---|---|---|
| Y6(-1,0,-1) | 0.52 | 0.52 | 0.52 | 0.52 | 0.51 | 1 |
| Y5(-1,0,-1) | 0.52 | 0.52 | 0.52 | 0.52 | 1 | 0.51 |
| Y4(-1,0,-1) | 0.52 | 0.52 | 0.52 | 1 | 0.52 | 0.52 |
| Y3(-1,0,-1) | 0.52 | 0.52 | 1 | 0.52 | 0.52 | 0.52 |
| Y2(-1) | 0.52 | 1 | 0.52 | 0.52 | 0.52 | 0.52 |
| Y1(.) | 1 | 0.52 | 0.52 | 0.52 | 0.52 | 0.52 |

**Web Figure 13:** Visualization of Area Under the Curve (AUC; panel a) and an approximation of AUC using the trapezoidal rule (b).

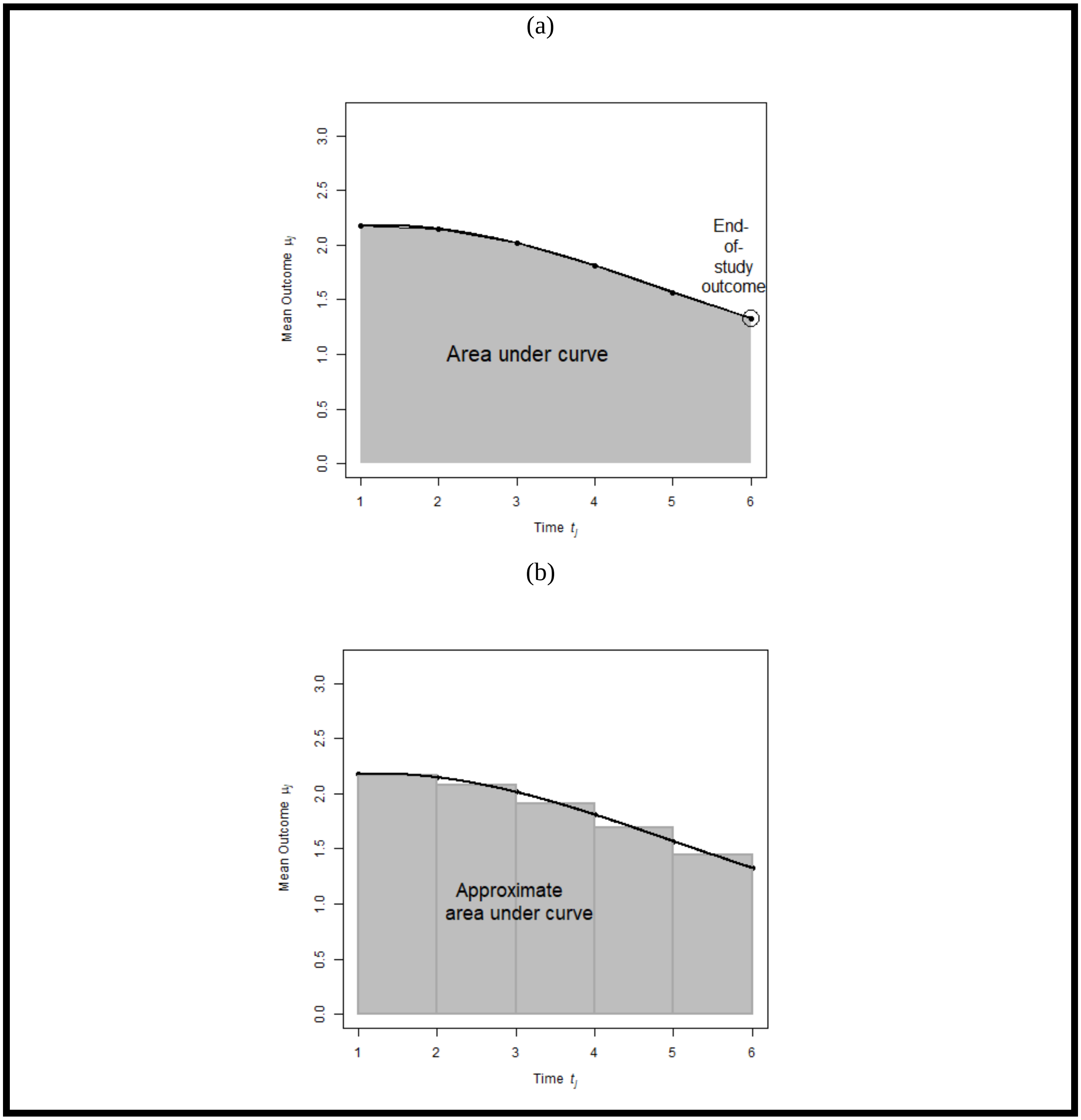

## Web Tables

(Web Tables begin on the following page)

**Web Table 1:** Rows display the mean trajectory of the longitudinal count outcome under each of the four EDTRs in terms of the parameters in Equation 1 and 2 of the main manuscript when $K = 2$ and $T = 6$, and $g$ is the log-link function.

| EDTR | $\mu_1^{(a_1,a_2^{NR})}(\boldsymbol{X})$ | $\mu_2^{(a_1,a_2^{NR})}(\boldsymbol{X})$ | $\mu_3^{(a_1,a_2^{NR})}(\boldsymbol{X})$ | $\mu_4^{(a_1,a_2^{NR})}(\boldsymbol{X})$ | $\mu_5^{(a_1,a_2^{NR})}(\boldsymbol{X})$ | $\mu_6^{(a_1,a_2^{NR})}(\boldsymbol{X})$ |
|---|---|---|---|---|---|---|
| $(+1,+1)$ | $exp(\boldsymbol{\beta}_X^T\boldsymbol{X} + \beta_{z,0})$ | $exp(\boldsymbol{\beta}_X^T\boldsymbol{X} + \beta_{z,0} + \beta_{z,1,2})$ | $exp(\boldsymbol{\beta}_X^T\boldsymbol{X} + \beta_{z,0} + \beta_{z,3,3})$ | $exp(\boldsymbol{\beta}_X^T\boldsymbol{X} + \beta_{z,0} + \beta_{z,3,4})$ | $exp(\boldsymbol{\beta}_X^T\boldsymbol{X} + \beta_{z,0} + \beta_{z,3,5})$ | $exp(\boldsymbol{\beta}_X^T\boldsymbol{X} + \beta_{z,0} + \beta_{z,3,6})$ |
| $(+1,-1)$ | $exp(\boldsymbol{\beta}_X^T\boldsymbol{X} + \beta_{z,0})$ | $exp(\boldsymbol{\beta}_X^T\boldsymbol{X} + \beta_{z,0} + \beta_{z,1,2})$ | $exp(\boldsymbol{\beta}_X^T\boldsymbol{X} + \beta_{z,0} + \beta_{z,4,3})$ | $exp(\boldsymbol{\beta}_X^T\boldsymbol{X} + \beta_{z,0} + \beta_{z,4,4})$ | $exp(\boldsymbol{\beta}_X^T\boldsymbol{X} + \beta_{z,0} + \beta_{z,4,5})$ | $exp(\boldsymbol{\beta}_X^T\boldsymbol{X} + \beta_{z,0} + \beta_{z,4,6})$ |
| $(-1,+1)$ | $exp(\boldsymbol{\beta}_X^T\boldsymbol{X} + \beta_{z,0})$ | $exp(\boldsymbol{\beta}_X^T\boldsymbol{X} + \beta_{z,0} + \beta_{z,2,2})$ | $exp(\boldsymbol{\beta}_X^T\boldsymbol{X} + \beta_{z,0} + \beta_{z,5,3})$ | $exp(\boldsymbol{\beta}_X^T\boldsymbol{X} + \beta_{z,0} + \beta_{z,5,4})$ | $exp(\boldsymbol{\beta}_X^T\boldsymbol{X} + \beta_{z,0} + \beta_{z,5,5})$ | $exp(\boldsymbol{\beta}_X^T\boldsymbol{X} + \beta_{z,0} + \beta_{z,5,6})$ |
| $(-1,-1)$ | $exp(\boldsymbol{\beta}_X^T\boldsymbol{X} + \beta_{z,0})$ | $exp(\boldsymbol{\beta}_X^T\boldsymbol{X} + \beta_{z,0} + \beta_{z,2,2})$ | $exp(\boldsymbol{\beta}_X^T\boldsymbol{X} + \beta_{z,0} + \beta_{z,6,3})$ | $exp(\boldsymbol{\beta}_X^T\boldsymbol{X} + \beta_{z,0} + \beta_{z,6,4})$ | $exp(\boldsymbol{\beta}_X^T\boldsymbol{X} + \beta_{z,0} + \beta_{z,6,5})$ | $exp(\boldsymbol{\beta}_X^T\boldsymbol{X} + \beta_{z,0} + \beta_{z,6,6})$ |

**Web Table 2:** The first row of the table below enumerates all possible potential outcomes corresponding to the simplified SMART design in Figure 1 of the main manuscript when there are three measurement occasions. Below, the potential outcomes $Y_{i,j}^{s}$'s that would be feasible for an individual, contingent on their subgroup membership, are denoted by a checkmark (✓).

| | $Y_{i,1}^{(\cdot)}$ | $Y_{i,2}^{(+1)}$ | $Y_{i,2}^{(-1)}$ | $Y_{i,3}^{(+1,1,0)}$ | $Y_{i,3}^{(+1,0,+1)}$ | $Y_{i,3}^{(+1,0,+1)}$ | $Y_{i,3}^{(-1,1,0)}$ | $Y_{i,3}^{(-1,0,+1)}$ | $Y_{i,3}^{(-1,0,-1)}$ |
|---|---|---|---|---|---|---|---|---|---|
| Subgroup 1 | ✓ | ✓ | ✓ | ✓ | − | − | ✓ | − | − |
| Subgroup 2 | ✓ | ✓ | ✓ | ✓ | − | − | − | ✓ | ✓ |
| Subgroup 3 | ✓ | ✓ | ✓ | − | ✓ | ✓ | ✓ | − | − |
| Subgroup 4 | ✓ | ✓ | ✓ | − | ✓ | ✓ | − | ✓ | ✓ |

**Web Table 3:** Constraints on the values of $Y_{i,2}^{(+1)}$ and $Y_{i,2}^{(-1)}$ for the four subgroups in a SMART are listed; these constraints are based on response status defined as $R_i^{(a_1)} = I\left(Y_{i,K}^{(a_1)} \leq c\right)$ in the proposed approach to data generation. Here, $K = 2$.

| | Constraint on the value of $Y_{i,2}^{(+1)}$ | Constraint on the value of $Y_{i,2}^{(-1)}$ |
|---|---|---|
| Subgroup 1 | $Y_{i,2}^{(+1)} \leq c$ | $Y_{i,2}^{(-1)} \leq c$ |
| Subgroup 2 | $Y_{i,2}^{(+1)} \leq c$ | $Y_{i,2}^{(-1)} > c$ |
| Subgroup 3 | $Y_{i,2}^{(+1)} > c$ | $Y_{i,2}^{(-1)} \leq c$ |
| Subgroup 4 | $Y_{i,2}^{(+1)} > c$ | $Y_{i,2}^{(-1)} > c$ |

**Web Table 4:** The complete specification of the marginal distribution of the components of $\boldsymbol{\theta}_i^{g}$ for each subgroup is displayed. Below, c is a cutoff used in the definition of response status. The CDF of a negative binomial (NB), upper truncated negative binomial (UTNB), and lower truncated negative binomial (LTNB) distribution determined by c, $\mu_j^{s}$, and $\zeta_j^{s}$ is denoted by $F_{NB\left(\mu_j^{s},\zeta_j^{s}\right)}$, $F_{UpperTruncNB\left(\mu_j^{s},\zeta_j^{s},c\right)}$, and $F_{LowerTruncNB\left(\mu_j^{s},\zeta_j^{s},c\right)}$, respectively. In terms of the probability mass function (PMF) of a NB random variable $f_{NB\left(\mu_j^{s},\zeta_j^{s}\right)}$, the PMF of a UTNB random variable and a LTNB random variable is $f_{UpperTruncNB\left(\mu_j^{s},\zeta_j^{s},c\right)}(w) = \frac{f_{NB\left(\mu_j^{s},\zeta_j^{s}\right)}(w)I(w\leq c)}{\sum_{y=0}^{c} f_{NB\left(\mu_j^{s},\zeta_j^{s}\right)}(y)}$ and $f_{LowerTruncNB\left(\mu_j^{s},\zeta_j^{s},c\right)}(w) = \frac{f_{NB\left(\mu_j^{s},\zeta_j^{s}\right)}(w)I(w>c)}{1-\sum_{y=0}^{c} f_{NB\left(\mu_j^{s},\zeta_j^{s}\right)}(y)}$, respectively. When c=0, the UTNB PMF reduces to a point mass at zero.

| | $F_{Y_{i,1}}$ | $F_{Y_{i,2}^{(+1)}}$ | $F_{Y_{i,2}^{(-1)}}$ |
|---|---|---|---|
| Subgroup 1 | $F_{NB\left(\mu_1^{(\cdot)},\zeta_1^{(\cdot)}\right)}$ | $F_{UpperTruncNB\left(\mu_2^{(+1)},\zeta_2^{(+1)},c\right)}$ | $F_{UpperTruncNB\left(\mu_2^{(-1)},\zeta_2^{(-1)},c\right)}$ |
| Subgroup 2 | $F_{NB\left(\mu_1^{(\cdot)},\zeta_1^{(\cdot)}\right)}$ | $F_{UpperTruncNB\left(\mu_2^{(+1)},\zeta_2^{(+1)},c\right)}$ | $F_{LowerTruncNB\left(\mu_2^{(-1)},\zeta_2^{(-1)},c\right)}$ |
| Subgroup 3 | $F_{NB\left(\mu_1^{(\cdot)},\zeta_1^{(\cdot)}\right)}$ | $F_{LowerTruncNB\left(\mu_2^{(+1)},\zeta_2^{(+1)},c\right)}$ | $F_{UpperTruncNB\left(\mu_2^{(-1)},\zeta_2^{(-1)},c\right)}$ |
| Subgroup 4 | $F_{NB\left(\mu_1^{(\cdot)},\zeta_1^{(\cdot)}\right)}$ | $F_{LowerTruncNB\left(\mu_2^{(+1)},\zeta_2^{(+1)},c\right)}$ | $F_{LowerTruncNB\left(\mu_2^{(-1)},\zeta_2^{(-1)},c\right)}$ |

**Web Table 5:** Web Table 4, Continued

| | $F_{Y_{i,3}^{(+1,1,0)}}$ | $F_{Y_{i,3}^{(+1,0,+1)}}$ | $F_{Y_{i,3}^{(+1,0,-1)}}$ | $F_{Y_{i,3}^{(-1,1,0)}}$ | $F_{Y_{i,3}^{(-1,0,+1)}}$ | $F_{Y_{i,3}^{(+1,0,-1)}}$ |
|---|---|---|---|---|---|---|
| Subgroup 1 | $F_{NB}\left(\mu_3^{(+1,1,0)}, \zeta_3^{(+1,1,0)}\right)$ | – | – | $F_{NB}\left(\mu_3^{(-1,1,0)}, \zeta_3^{(-1,1,0)}\right)$ | – | – |
| Subgroup 2 | $F_{NB}\left(\mu_3^{(+1,1,0)}, \zeta_3^{(+1,1,0)}\right)$ | – | – | – | $F_{NB}\left(\mu_3^{(-1,0,+1)}, \zeta_3^{(-1,0,+1)}\right)$ | $F_{NB}\left(\mu_3^{(-1,0,-1)}, \zeta_3^{(-1,0,-1)}\right)$ |
| Subgroup 3 | – | $F_{NB}\left(\mu_3^{(+1,0,+1)}, \zeta_3^{(+1,0,+1)}\right)$ | $F_{NB}\left(\mu_3^{(+1,0,-1)}, \zeta_3^{(+1,0,-1)}\right)$ | $F_{NB}\left(\mu_3^{(-1,1,0)}, \zeta_3^{(-1,1,0)}\right)$ | – | – |
| Subgroup 4 | – | $F_{NB}\left(\mu_3^{(+1,0,+1)},\zeta_3^{(+1,0,+1)}\right)$ | $F_{NB}\left(\mu_3^{(+1,0,-1)}, \zeta_3^{(+1,0,-1)}\right)$ | – | $F_{NB}\left(\mu_3^{(-1,0,+1)}, \zeta_3^{(-1,0,+1)}\right)$ | $F_{NB}\left(\mu_3^{(-1,0,-1)}, \zeta_3^{(-1,0,-1)}\right)$ |

**Web Table 6:** Parameter values in Simulation Study 3. Below, $\pi_j^s$ denotes $Pr\{Y_j^s = 0\}$.

| ***Fixed across all scenarios:*** | |
|---|---|
| ***Total Sample Size*** | ***ETS Means*** |
| N=100, 150, 200, …, 550 | For all s: $\mu_1^s = 2.5, \mu_2^s = 4.8, \mu_3^s = 2.6, \mu_4^s = 2.7, \mu_5^s = 2.75, \mu_6^s = 2.8$ |
| ***Copula Dependence Parameter*** | |
| $\rho = 0.2, 0.4, 0.6$ | |
| ***Proportion of Responders*** | |
| $p = q = 0.40$ | |

| ***Varied across scenarios:*** | | |
|---|---|---|
| ***Scenario 1*** | ***Scenario 2*** | ***Scenario 3*** |
| ***ETS Proportion of Zeros*** | ***ETS Proportion of Zeros*** | ***ETS Proportion of Zeros*** |
| $\pi_1^{(\cdot)} = 0.40, \pi_2^{(+1)} = 0.40, \pi_2^{(-1)} = 0.20$ | $\pi_1^{(\cdot)} = 0.40, \pi_2^{(+1)} = 0.40, \pi_2^{(-1)} = 0.40$ | $\pi_1^{(\cdot)} = 0.40, \pi_2^{(+1)} = 0.40, \pi_2^{(-1)} = 0.60$ |
| $\pi_3^{(+1,0,0)} = \pi_3^{(+1,0,+1)} = \pi_3^{(+1,0,-1)} = 0.20$ | $\pi_3^{(+1,0,0)} = \pi_3^{(+1,0,+1)} = \pi_3^{(+1,0,-1)} = 0.40$ | $\pi_3^{(+1,0,0)} = \pi_3^{(+1,0,+1)} = \pi_3^{(+1,0,-1)} = 0.60$ |
| $\pi_4^{(+1,0,0)} = \pi_4^{(+1,0,+1)} = \pi_4^{(+1,0,-1)} = 0.20$ | $\pi_4^{(+1,0,0)} = \pi_4^{(+1,0,+1)} = \pi_4^{(+1,0,-1)} = 0.40$ | $\pi_4^{(+1,0,0)} = \pi_4^{(+1,0,+1)} = \pi_4^{(+1,0,-1)} = 0.60$ |
| $\pi_5^{(+1,0,0)} = \pi_5^{(+1,0,+1)} = \pi_5^{(+1,0,-1)} = 0.20$ | $\pi_5^{(+1,0,0)} = \pi_5^{(+1,0,+1)} = \pi_5^{(+1,0,-1)} = 0.40$ | $\pi_5^{(+1,0,0)} = \pi_5^{(+1,0,+1)} = \pi_5^{(+1,0,-1)} = 0.60$ |
| $\pi_6^{(+1,0,0)} = \pi_6^{(+1,0,+1)} = \pi_6^{(+1,0,-1)} = 0.20$ | $\pi_6^{(+1,0,0)} = \pi_6^{(+1,0,+1)} = \pi_6^{(+1,0,-1)} = 0.40$ | $\pi_6^{(+1,0,0)} = \pi_6^{(+1,0,+1)} = \pi_6^{(+1,0,-1)} = 0.60$ |
| $\pi_3^{(-1,0,0)} = \pi_3^{(-1,0,+1)} = \pi_3^{(-1,0,-1)} = 0.20$ | $\pi_3^{(-1,0,0)} = \pi_3^{(-1,0,+1)} = \pi_3^{(-1,0,-1)} = 0.40$ | $\pi_3^{(-1,0,0)} = \pi_3^{(-1,0,+1)} = \pi_3^{(-1,0,-1)} = 0.60$ |
| $\pi_4^{(-1,0,0)} = \pi_4^{(-1,0,+1)} = \pi_4^{(-1,0,-1)} = 0.20$ | $\pi_4^{(-1,0,0)} = \pi_4^{(-1,0,+1)} = \pi_4^{(-1,0,-1)} = 0.40$ | $\pi_4^{(-1,0,0)} = \pi_4^{(-1,0,+1)} = \pi_4^{(-1,0,-1)} = 0.60$ |
| $\pi_5^{(-1,0,0)} = \pi_5^{(-1,0,+1)} = \pi_5^{(-1,0,-1)} = 0.20$ | $\pi_5^{(-1,0,0)} = \pi_5^{(-1,0,+1)} = \pi_5^{(-1,0,-1)} = 0.40$ | $\pi_5^{(-1,0,0)} = \pi_5^{(-1,0,+1)} = \pi_5^{(-1,0,-1)} = 0.60$ |
| $\pi_6^{(-1,0,0)} = \pi_6^{(-1,0,+1)} = \pi_6^{(-1,0,-1)} = 0.20$ | $\pi_6^{(-1,0,0)} = \pi_6^{(-1,0,+1)} = \pi_6^{(-1,0,-1)} = 0.40$ | $\pi_6^{(-1,0,0)} = \pi_6^{(-1,0,+1)} = \pi_6^{(-1,0,-1)} = 0.60$ |

| ***Value of dispersion parameter $\zeta_j^s$ implied by choice of values for $\mu_j^s$ and $\pi_j^s$:*** | | |
|---|---|---|
| ***Scenario 1*** | ***Scenario 2*** | ***Scenario 3*** |
| For all s: $\zeta_1^s = 0.51, \zeta_2^s = 1.18, \zeta_3^s = 0.55, \zeta_4^s = 0.60, \zeta_5^s = 0.62, \zeta_6^s = 0.63$. | For all s: $\zeta_1^s = 1.92, \zeta_2^s = 2.98, \zeta_3^s = 1.98, \zeta_4^s = 2.05, \zeta_5^s = 2.08, \zeta_6^s = 2.11$. | For all s: $\zeta_1^s = 5.15, \zeta_2^s = 6.91, \zeta_3^s = 5.26, \zeta_4^s = 5.36, \zeta_5^s = 5.41, \zeta_6^s = 5.46$. |